\documentclass[superscriptaddress,amsmath,amssymb,
aps,prd,notitlepage,longbibliography,floatfix,nofootinbib]{revtex4-1}

\usepackage{tensor}     
\usepackage{graphicx}   
\usepackage{subfigure}
\usepackage[
colorlinks=true,        
citecolor=blue,         
linkcolor=blue,         
urlcolor=blue           
]{hyperref}             
\usepackage{bm}         
\usepackage{xcolor}     
\usepackage{lipsum}
\usepackage{color}      
\usepackage[utf8]{inputenc} 
\usepackage[section]{placeins} 
\usepackage{multirow}
\usepackage[title]{appendix}
\newcommand{\nc}{\newcommand*}

\nc{\al}{\alpha}
\nc{\s}{\sigma}
\nc{\kp}{\kappa}
\nc{\dt}{\delta}
\nc{\Dt}{\Delta}
\nc{\Ld}{\Lambda}
\nc{\p}{\partial}
\nc{\Gm}{\Gamma}
\nc{\om}{\omega}
\nc{\Om}{\Omega}
\nc{\rd}{\mathrm{d}}
\def\({\left(}
\def\){\right)}
\def\[{\left[}
\def\]{\right]}
\def\e{\begin{equation}}
\def\q{\end{equation}}
\def\m{\begin{eqnarray}}
\def\n{\end{eqnarray}}
\nc{\Eq}[1]{Eq.~\eqref{#1}}     
\nc{\Fig}[1]{Fig.~\ref{#1}}     
\nc{\Table}[1]{Table~\ref{#1}}  
\nc{\Sec}[1]{Sec.~\ref{#1}}     
\nc{\Msun}{M_\odot}             
\nc{\fpbh}{f_{\mathrm{pbh}}}    
\nc{\fpbhn}{f_{\mathrm{pbh0}}}    
\nc{\mR}{\mathcal{R}} 
\nc{\seq}{\sigma_{\mathrm{eq}}}
\nc{\ogw}{\Omega_{\mathrm{GW}}}
\nc{\gpcyr}{\mathrm{Gpc}^{-3}\,\mathrm{yr}^{-1}}
\nc{\lvc}{LIGO/Virgo} 
\nc{\SNR}{\mathrm{SNR}} 
\nc{\mmin}{{m_{\mathrm{min}}}}
\nc{\mmax}{{m_{\mathrm{max}}}}
\nc{\Mmin}{{M_{\mathrm{min}}}}
\nc{\fmin}{{f_{\mathrm{min}}}}
\nc{\VT}{\mathrm{VT}}
\nc{\rhoGW}{\rho_{\mathrm{GW}}}
\nc{\vth}{\vec{\theta}}
\nc{\vd}{\vec{d}}
\nc{\vla}{\vec{\lambda}}
\nc{\Nobs}{N_{\mathrm{obs}}}
\nc{\av}[1]{\langle #1 \rangle} 
\nc{\km}{\mathrm{km}}
\nc{\Mpc}{\mathrm{Mpc}}
\nc{\Tobs}{T_{\mathrm{obs}}}
\nc{\Ntemp}{N_{\mathrm{temp}}}
\nc{\fyr}{f_{\mathrm{yr}}}
\nc{\addref}{[\textcolor{red}{add ref}] } 
\nc{\eg}{\textit{e.g.~}}
\nc{\app}{\approx}
\nc{\hf}{\frac{1}{2}}
\nc{\discuss}{\textcolor{red}{Add discussion here!}}
\nc{\red}[1]{\textcolor{red}{#1}}
\nc{\hp}{h_+} 
\nc{\hc}{h_{\times}} 
\nc{\Oh}{\hat{\Omega}}
\nc{\vx}{\vec{x}}
\nc{\mh}{\hat{m}}
\nc{\nh}{\hat{n}}
\nc{\zh}{\hat{z}}
\nc{\ph}{\hat{p}}
\nc{\A}[1]{\mathcal{A}_{#1}}
\nc{\Ogw}[1]{\Omega_{\mathrm{#1}}}
\nc{\bn}[1]{\dt\bm{t}_{\text{#1}}}
\nc{\bC}[1]{\bm{C}_{\text{#1}}}
\nc{\NTOA}{N_{\text{TOA}}}
\nc{\Nmode}{{N_{\text{mode}}}}
\nc{\ARN}{A_{\rm{RN}}}
\nc{\gRN}{\gamma_{\rm{RN}}}
\nc{\bS}{\mathbf{\Sigma}}
\nc{\br}{\mathbf{r}}
\nc{\bN}{\mathbf{R}}
\nc{\Agw}{A_\mathrm{GWB}}
\nc{\UCP}{\mathrm{UCP}}
\nc{\TT}{\mathrm{TT}}
\nc{\ST}{\mathrm{ST}}
\nc{\SL}{\mathrm{SL}}
\nc{\VL}{\mathrm{VL}}

\nc{\BFST}{$107 \pm 7$}

\begin{document}
	
\title{On the Interaction between Ultralight Bosons and Quantum-Corrected Black Holes}
	

\author{Rong-Zhen Guo}
\affiliation{CAS Key Laboratory of Theoretical Physics,
	Institute of Theoretical Physics, Chinese Academy of Sciences,
	Beijing 100190, China}
\affiliation{School of Physical Sciences,
	University of Chinese Academy of Sciences,
	No. 19A Yuquan Road, Beijing 100049, China}
\author{Chen Yuan}
\affiliation{CAS Key Laboratory of Theoretical Physics,
Institute of Theoretical Physics, Chinese Academy of Sciences,
Beijing 100190, China}
\affiliation{School of Physical Sciences,
University of Chinese Academy of Sciences,
No. 19A Yuquan Road, Beijing 100049, China}

\author{Qing-Guo Huang}
\email{Corresponding author: huangqg@itp.ac.cn}
\affiliation{CAS Key Laboratory of Theoretical Physics,
Institute of Theoretical Physics, Chinese Academy of Sciences,
Beijing 100190, China}
\affiliation{School of Physical Sciences,
University of Chinese Academy of Sciences,
No. 19A Yuquan Road, Beijing 100049, China}
\affiliation{School of Fundamental Physics and Mathematical Sciences
Hangzhou Institute for Advanced Study, UCAS, Hangzhou 310024, China}

\date{\today}

\begin{abstract}

Both ultralight dark matter and exploring the quantum nature of black holes are all topics of great interest in gravitational wave astronomy at present. The superradiant instability allows an exotic compact object (ECO) to be surrounded by an ultralight boson cloud, which leads to the emission of gravitational waves and further triggers rich dynamical effects. 
In this paper, we study the gravitational effects of superradiant instabilities by calculating the energy fluxes of gravitational waves emitted from ultralight scalar dark matter fields by solving the Teukolsky equation in the background of a massive ECO phenomenologically described by a Kerr geometry with a reflective boundary condition at its physical boundary. We find that both the amplitude and phase of the reflectivity will either suppress or enhance the energy flux of GWs by several orders of magnitude if $M\mu \gtrsim 0.5$ where $M$ and $\mu$ are the mass of ECO and boson, respectively. However, the modifications to energy flux are negligible if $M \mu \lesssim 0.5$. Our results suggest that reflectivity will play a significant role in the near-horizon physics of ECO.
\end{abstract}

\maketitle

\maketitle
\section{Introduction}

Ultralight bosonic particles are predicted in several beyond Standard Model theory \cite{Arvanitaki:2009fg,Essig:2013lka,Irastorza:2018dyq,Goodsell:2009xc,Jaeckel:2010ni,Antypas:2022asj}, which can be one of possible dark matter candidates\cite{Preskill:1982cy,Abbott:1982af,Dine:1982ah,svrcek2006axions,Arvanitaki:2009fg,Arvanitaki:2010sy,Essig:2013lka,Brito:2015oca,Marsh:2015xka,Hui:2016ltb,Annulli:2020lyc,Chadha-Day:2021szb,Davoudiasl:2020uig,Antypas:2022asj,Adams:2022pbo}. Detecting ultralight dark matter (UDM) is rather difficult by traditional particle physics experiments, while fortunately, various phenomena caused by UDM in astrophysics can help us to alleviate this dilemma. For instance, using electromagnetic waves (e.g., the Event Horizon Telescope observations of M87*) \cite{Davoudiasl:2019nlo,Chen:2021lvo} and gravitational waves (GWs) \cite{Damour:1976kh,Zouros:1979iw,Detweiler:1980uk,Dolan,string_axiverse,Shlapentokh-Rothman:2013ysa,Pani:2012vp,Pani:2012bp,Witek:2012tr,sr_tensor,Endlich:2016jgc,East:2017mrj,East:2017ovw,sr_vector_4,Cardoso:2018tly,East:2018glu,Frolov:2018ezx,Dolan:2018dqv,Baumann:2019eav,Brito:2020lup,Tsukada:2020lgt,Yuan:2021ebu,Davoudiasl:2021ijv,Chung:2021roh,PPTA:2022eul,Yuan:2022bem} to probe ultralight bosons has become an active emerging field. 

The mechanism responsible for the astrophysical implication is the superradiant instabilities in a UDM-spinning black hole (BH) system (see \cite{Brito:2015oca}, a review on superradiance). When the Compton wavelength of bosonic UDM is close to the same order as the BH radius, the energy and angular momentum of the BH would be transformed into the bosonic field, forming a macroscopic co-rotating boson cloud, which dissipates its energy through the emission of nearly monochromatic GWs. Therefore, the boson-BH system can act as a continuous GW source and is detectable by GW observations \cite{Arvanitaki:2014wva,Arvanitaki:2016qwi,Baryakhtar:2017ngi,Brito:2017wnc,Brito:2017zvb,Isi:2018pzk,Ghosh:2018gaw,Palomba:2019vxe,Sun:2019mqb,Zhu:2020tht,Brito:2020lup,Ng:2020jqd,Tsukada:2020lgt,Yuan:2021ebu,Yuan:2022bem}. Until now, the null detection of GWs from the boson-BH systems has constrained bosonic UDM in various mass windows.  \cite{Kodama:2011zc,Brito:2014wla,Arvanitaki:2010sy,Brito:2017wnc,Brito:2017zvb,Tsukada:2018mbp,Zhu:2020tht,Tsukada:2020lgt,Chen:2021lvo,Yuan:2022bem}.

In addition to the dark matter puzzle, another ``holy grail" of theoretical physics is understanding the nature of BHs. In general relativity, a BH is characterized by the presence of its smooth one-way membrane-like event horizon \cite{Thorne:1986iy}, which encloses the essential singularity predicted by the Penrose-Hawking singularity theorem \cite{Hawking:1973uf}, and the Cosmic Censorship Conjecture protects the essential singularity inside from being seen by observers outside the event horizon. However, BH information paradox \cite{polchinski2017black} and the thirst for quantum gravity drive the innovation of this traditional paradigm. Several high-profile candidates of quantum gravity, such as String Theory and Loop Quantum Gravity, predict different quantum-corrected BH models (e.g. fuzzball \cite{skenderis2008fuzzball} in String Theory and black to white hole transition in Loop Quantum Gravity \cite{Haggard:2014rza,Bianchi:2018mml}). Besides, firewall \cite{almheiri2013black} and area quantization (named Bekenstein-Mukhanov ansatz) \cite{Bekenstein:1974jk,Bekenstein:1995ju} could also challenge the traditional paradigm. 
However, the longstanding lack of observational prediction of these dazzling quantum-corrected BH models obstructs our understanding of the level of phenomenology. How to test them via observation becomes a pressing issue \cite{Zurek:2022xzl,deBoer:2022zka}. And studying their dynamical modifications to classical BHs through GWs opens a new window to fill the gap \cite{Cardoso:2016rao,Cardoso:2019apo,mark2017recipe,Price:2017cjr,Fang:2021iyf}. Here we use the model that quantum corrections of BHs can be described by modifying the boundary conditions of various types of test fields at their physical boundaries on classical BHs. The effectiveness of this model establishes the commonality of amounts of these quantum-corrected BH models (though not all) that their outer geometries are close to classical GR solutions. Thus, their modifications of gravitational dynamics in the sense of perturbation towards classical BHs can be described by a reflectivity $\mathbf{R}(\omega)$ near the physical boundary (we refer to it as ``would-be horizon''), which lead to ``echoes'' of GWs. Indeed, the reflective nature of several models, e.g., wormhole \cite{Biswas:2022wah,Bueno:2017hyj}, area quantization \cite{Cardoso:2019apo}, and fuzzball \cite{Ikeda:2021uvc,Abedi:2016hgu}, have been proved. Besides, one could also regard reflectivity as a conclusion of Point Particle Effective Field Theory \cite{Rummel:2019ads,Burgess:2018pmm}. 

The reflective features of these quantum-corrected BH models imply that the nature of BHs can be studied by reanalyzing those physical systems described by BH perturbation theory \cite{brito2015superradiance}, such as quasinormal modes \cite{McManus:2020lgm,Ikeda:2021uvc}, ringdown waveform \cite{Ou:2021efv,Mark:2017dnq}, extreme mass ratio inspirals \cite{Fang:2021iyf,Maggio:2021uge} and tidal heating \cite{Datta:2019epe}. Superradiant instabilities are thus no exception. Formation of the UDM cloud,  emission of GWs, and dynamics caused by the nonlinear interaction of UDM or binary background have been studied thoroughly in \cite{Dolan:2007mj,Brito:2014wla,East:2017mrj,East:2018glu,Baryakhtar:2020gao,Yoshino:2013ofa,Siemonsen:2019ebd,East:2017ovw}. And these researches have been used to constrain UDM \cite{Yuan:2021ebu,Brito:2017wnc,Brito:2017zvb,Isi:2018pzk,Sun:2019mqb,Ng:2020jqd}.  For the quantum-corrected BH models, we obtained analytical results in our previous work that reflectivity can shift both energy levels and the characteristic frequencies of the bosonic cloud \cite{Guo:2021xao}. This paper aims to continue our former work -- investigating the GW emission by a scalar UDM cloud on a quantum-corrected BH.

In this paper, we calculate the energy fluxes of GWs emitted from UDM by numerically solving the Teukolsky equation in the background of a quantum-corrected BH described phenomenologically by a Kerr-like exotic compact object (ECO) with a reflective boundary condition at its ``would-be horizon''. Our results show that once the bosonic cloud is in the region close to the ``would-be horizon'' horizon, the energy flux from the whole system will change dramatically compared to the BH. The energy flux is sensitive to the amplitude and phase correction of the reflectivity. And our results also show a complex and interesting dependence on all related parameters for energy flux, such as the ratio of bosonic Compton wavelength and Schwarzschild radius and spin of the ECO.

The paper is organized as follows. In Sec.~II, we briefly review the superradiant instabilities and the formation of a bosonic cloud around a spinning ECO. In Sec.~III, we introduce the Teukolsky formula and its application to calculate the energy flux of a given source for a Kerr-like ECO. Both the source term and Green's function are constructed. And here the reflectivity is introduced in the Sasaki-Nakamura equation so that it can have a physical reflecting meaning in the view of the standard wave equation. Our numerical results are shown in In Sec.~IV. Both constant reflectivity and other physically motivated models are studied. Finally, we summarize our results in Sec.~V. Besides, in Appendix A, we introduce Leaver’s continued fractions method to numerically construct the homogeneous solutions to Teukolsky equations to be used in the main text. And in Appendix B, we compare the difference between energy fluxes calculated by the metric reconstruction method and the Teukolsky method, and it turns out that the approach directly based on the Teukolsky equation is more plausible. Throughout the paper, we use $c=G=\hbar=1$ and the signature $(-1,+1,+1,+1)$.

\section{\label{section1} Superradiant Instabilities and Bosonic Clouds of Spinning Exotic Compact Objects}
We focus on the ECO whose modification to a Kerr BH is only significant near its event horizon so that the external geometry could be well-described by the Kerr metric in the Boyer-Lindquist coordinates $(t, r, \theta, \phi)$ such that
\begin{equation}
\begin{aligned}
&d s^{2}=-\frac{\Delta-a^{2} \sin ^{2} \theta}{\Sigma} d t^{2}-\frac{4 M a r \sin ^{2} \theta}{\Sigma} d t d \varphi \\
&+\left[\frac{\left(r^{2}+a^{2}\right)^{2}-\Delta a^{2} \sin ^{2} \theta}{\Sigma}\right] \sin ^{2} \theta d \varphi^{2}+\frac{\Sigma}{\Delta} d r^{2}+\Sigma d \theta^{2},
\end{aligned}
\end{equation}
with
\begin{equation}
\Sigma=r^{2}+a^{2} \cos ^{2} \theta, \quad \Delta=r^{2}+a^{2}-2 M r,
\end{equation}
and $r_{\pm}=M\pm\sqrt{M^2-a^2}$ is the location of the event horizon and inner horizon respectively. In such a case, the field falling into the ECO will have a certain probability to bounce back near its physical boundary. The dynamics of a field on a given Kerr-like ECO background should be governed by the same differential equation on the Kerr background except for its boundary condition near the event horizon. We consider a real scalar field $\Phi$ on the Kerr-like ECO background with its self-interaction and  gravitational back-reaction neglected \cite{Yoshino:2013ofa}. So $\Phi$ satisfies massive Klein-Gordon equation :
\begin{equation}
\nabla^{2} \Phi-\mu^{2} \Phi=0, 
\end{equation}
where $\mu$ stands for the mass of $\Phi$.

In the external spacetime of Kerr-like ECO, $\Phi$ can be expressed as: 
\begin{equation}
    \Phi=\Re\left[\sum_{lm}\int_{-\infty}^{\infty}e^{-i \omega t+i m \phi} ~_{0}R_{\omega\ell m}(r) ~_{0}S_{\omega\ell m}(\theta) d\omega\right], 
\end{equation}
where $\Re$ stands for real parts and $\omega$ is the eigenfrequency. Here $\!_{0}R_{\omega\ell m}(r)$ and $\!_{0}S_{\omega\ell m}(r)$ satisfy
\begin{equation}
\begin{aligned}
&\frac{d}{d r}\left( \Delta \frac{d \,_{0}R_{\omega\ell m}}{d r}\right)+\left[\frac{K^{2}}{\Delta}-\lambda_{\ell m}-\mu^{2} r^{2}\right] \!_{0}R_{\omega\ell m}=0, \\
&\frac{1}{\sin \theta}\!\left(\!\frac{d}{d \theta}\!\sin\!\theta \frac{d\!\,_{0}S_{\omega\ell m}}{d \theta}\!\right)\!-\!\left[\!k^{2}\!a^{2}\!\cos ^{2} \!\theta\!+\!\frac{m^{2}}{\sin ^{2}\!\theta}\!-\!A_{\ell m}\right]\!_{0}\!S_{\omega\ell m}\!=\!0,\label{scalareq}
\end{aligned}
\end{equation}
and 
\begin{equation}
\begin{aligned}
    &k=\sqrt{\mu^{2}-\omega^{2}},\\
    &K=\left(r^{2}+a^{2}\right) \omega-a m,\\
    &\lambda_{\ell m}=A_{\ell m}+a^{2} \omega^{2}-2 a m \omega .
\end{aligned}
\end{equation}

The calculation for superradiant instabilities of a massive scalar field in Kerr spacetime \cite{Dolan:2007mj,Detweiler:1980uk,Brito:2015oca} is similar to calculating the eigenvalues of bound states of hydrogen atoms: select decaying boundary condition at infinity and physical boundary condition at another point such that it constitutes a boundary-value problem. So superradiant instabilities occur only for several discrete states with their characterized frequencies $\{\omega_n \}$. However, unlike the case of the wave function of a hydrogen atom, the boundary-value problem of a BH is not self-adjoint since it admits outer field escape through its event horizon. Its non-self-adjoint nature leads to complex eigenvalues, that is, quasi-bound states with $\{\omega_n=\omega_n^R+i\omega_n^I \}$, where the corresponding state exponentially decays for $\omega_n^I<0$ and grows for $\omega_n^I>0$. The latter is the state suffering from superradiant instabilities. Moreover, one can show that superradiant instabilities will occur when $\omega^{R}_n<m \Omega_{H}$, where $\Omega_{H}=a /\left(r_{+}^{2}+a^{2}\right)$ is the angular velocity of a Kerr BH. In the case of Kerr-like ECO, it has been shown from our previous work that a real scalar field in a Kerr-like ECO background will also have superradiant instabilities \cite{Guo:2021xao}, thus forming a bosonic cloud around the Kerr-like ECO \cite{Brito:2017zvb,Brito:2014wla}. Assuming quasi-adiabatic approximation, such that the scalar field is nearly stationary with its $\omega_n^R$ as ``energy level'' and $\tau_{\text {inst }} \equiv 1 / \omega_n^{I}$ as the characteristic timescale of superradiant instabilities. Then the evolution of the bosonic cloud could be directly described by the real part of unstable quasi-bound state $\Phi_{nlm}$ with its time-dependent term neglected. See e.g., \cite{Yoshino:2013ofa,Brito:2017zvb}.

\section{Gravitational radiation from Kerr-like ECO-Boson System}

\subsection{Teukolsky Equation}
Once the bosonic cloud is given, one can calculate its GW emission and the GW radiation rate using BH perturbation theory \cite{Bardeen:1972fi,Press:1973zz,Teukolsky:1974yv,Sasaki:2003xr,Chandrasekhar:1985kt}. Test field in Kerr metric could be systematically described in the framework of Newman-Penrose formula. Given  null tetrads:
\begin{equation}
\begin{aligned}
l^{\mu} &=\frac{1}{\Delta}\left(r^{2}+a^{2}, 1,0, a\right), \\
n^{\mu} &=\frac{1}{2 \Sigma}\left(r^{2}+a^{2},-\Delta, 0, a\right), \\
m^{\mu} &=\frac{1}{\sqrt{2}(r+i a \cos \theta)}\left(i a \sin \theta, 0,1, \frac{i}{\sin \theta}\right),\label{nulltetrads}
\end{aligned}
\end{equation}
the outgoing gravitational radiation at null infinity is described by Newman-Penrose scalar $\psi_{4}$ with spin weight $s=-2$. The perturbation equation for $\phi \equiv \rho^{-4} \psi_{4}$, where $\rho=(r-i a \cos \theta)^{-1}$, is the well-known Teukolsky equation:
\begin{equation}
    { }_{s} \mathcal{O} \phi=4 \pi \Sigma \hat{T}.\label{TeukolskyOperator}
\end{equation}
Here ${ }_{s} \mathcal{O}$ reads:
\begin{equation}
\begin{aligned}
{ }_{s} \mathcal{O}&=-\left[\frac{\left(r^{2}+a^{2}\right)^{2}}{\Delta}-a^{2} \sin ^{2} \theta\right] \partial_{t}^{2}-\frac{4 M a r}{\Delta} \partial_{t} \partial_{\phi} \\
&-\left[\frac{a^{2}}{\Delta}-\frac{1}{\sin ^{2} \theta}\right] \partial_{\phi}^{2}+\Delta^{-s} \partial_{r}\left(\Delta^{s+1} \partial_{r}\right)+\frac{1}{\sin \theta} \partial_{\theta}\left(\sin \theta \partial_{\theta}\right) \\
&+2 s\left[\frac{a(r-M)}{\Delta}+\frac{i \cos \theta}{\sin ^{2} \theta}\right] \partial_{\phi}-s\left(s \cot ^{2} \theta-1\right)\\
&+2 s\left[\frac{M\left(r^{2}-a^{2}\right)}{\Delta}-r-i a \cos \theta\right] \partial_{t},
\end{aligned}
\end{equation}
and $\hat{T}$ is the source term of $\phi$.

Teukolsky equation is seperable according to the hidden symmetry of Kerr metric \cite{Frolov:2017kze}. Decomposing $\rho^{-4} \psi_{4}$ into:
\begin{equation}
\rho^{-4} \psi_{4}=\sum_{\tilde{\ell} \tilde{m}} \int d \tilde{\omega} e^{-i \tilde{\omega} t+i \tilde{m} \varphi}{ }_{-2} S_{\tilde{\ell} \tilde{m}}(\theta) R_{\tilde{\ell} \tilde{m} \tilde{\omega}}(r),
\end{equation}
(here we use tilde to distinguish $\psi_4$-related term from $\Phi$-related term) then Eq.~(\ref{TeukolskyOperator}) can be reduced to radial Teukolsky equation and angular Teukolsky equation such that:
\begin{equation}
\Delta^{2} \frac{d}{d r}\left(\frac{1}{\Delta} \frac{d R_{\tilde{\ell} \tilde{m} \tilde{\omega}}}{d r}\right)-V(r) R_{\tilde{\ell} \tilde{m} \tilde{\omega}}=T_{\tilde{\ell} \tilde{m} \tilde{\omega}},\label{TRE}
\end{equation}

\begin{equation}
	\left[\frac{1}{\sin \theta} \frac{d}{d \theta}\left(\sin \theta \frac{d}{d \theta}\right)-a^2 \tilde{\omega}^2 \sin ^2 \theta-\frac{(\tilde{m}-2 \cos \theta)^2}{\sin ^2 \theta}+4 a \tilde{\omega} \cos \theta-2+2 \tilde{m} a \tilde{\omega}+\tilde{\lambda}\right]{}_{-2} S_{\tilde{\ell} \tilde{m}}=0
\end{equation}
The potential $V(r)$ in Eq.~(\ref{TRE}) is given by
\begin{equation}
V(r)=-\frac{K^{2}+4 i(r-M) K}{\Delta}+8 i \tilde{\omega} r+\tilde{\lambda},
\end{equation}
where $K=\left(r^{2}+a^{2}\right) \tilde{\omega}-\tilde{m} a$, $\tilde{\lambda}$ is the eigenvalue of ${}_{-2} S_{\tilde{\ell} \tilde{m}}$, the spin-weighted spheroidal harmonic with $s=-2$, which is normalized by
\begin{equation}
\int_{0}^{\pi}\left| {}_{-2} S_{\tilde{\ell} \tilde{m}}\right|^{2} \sin \theta d \theta=1.
\end{equation}

\subsection{Source Term}
The energy-momentum tensor of massive scalar field is \cite{Yoshino:2013ofa}
\begin{equation}
T_{\mu \nu}(\Phi, \Phi)=\nabla_{\mu} \Phi \nabla_{\nu} \Phi-\frac{1}{2} g_{\mu \nu}\left(\nabla_{\rho} \Phi \nabla^{\rho} \Phi+\mu^{2} \Phi^2\right),\label{EMTensor}
\end{equation}
where $g_{\mu\nu}$ is Kerr metric, $\nabla_{\mu}$ is the covariant derivative compatiable with $g_{\mu\nu}$. For simplicity, here we only consider the single superradiant unstable mode,
\begin{equation}
    \Phi_{lm}=\Re \left[e^{-i \omega_n t+i m \phi} ~_{0}R_{\omega_n\ell m}(r) ~_{0}S_{\omega_n\ell m}(\theta)\right].
\end{equation}
Its corresponding source term in radial Teukolsky equation Eq.~(\ref{TRE}), $T_{\tilde{\ell} \tilde{m} \tilde{\omega}}$, can be expressed as \cite{Sasaki:2003xr}:
\begin{equation}
T_{\tilde{\ell} \tilde{m} \tilde{\omega}}=4 \int d \Omega d t \rho^{-5} \rho^{*^{-1}}\left(B_{2}^{\prime}+B_{2}^{\prime *}\right) e^{-i \tilde{m} \varphi+i \tilde{\omega} t} \frac{{}_{-2} S_{\tilde{\ell} \tilde{m}}}{\sqrt{2 \pi}},
\label{sourceTeukolsky}
\end{equation}
where the symbol ``$*$'' stands for complex conjugate, and
\begin{equation}
\begin{aligned}
B_{2}^{\prime} &=-\frac{1}{2} \rho^{8} \rho^* L_{-1}\left[\rho^{-4} L_{0}\left(\rho^{-2} \rho^{*^{-1}} T_{n n}\right)\right] \\
&-\frac{1}{2 \sqrt{2}} \rho^{8} \rho^* \Delta^{2} L_{-1}\left[\rho^{-4} \rho^{*^{2}} J_{+}\left(\rho^{-2} \rho^{*^{-2}} \Delta^{-1} T_{m^* n}\right)\right]; \\
B_{2}^{\prime *} &=-\frac{1}{4} \rho^{8} \rho^* \Delta^{2} J_{+}\left[\rho^{-4} J_{+}\left(\rho^{-2} \rho^* T_{m^* m^*}\right)\right] \\
&-\frac{1}{2 \sqrt{2}} \rho^{8} \rho^* \Delta^{2} J_{+}\left[\rho^{-4} \rho^{*^{2}} \Delta^{-1} L_{-1}\left(\rho^{-2} \rho^{*^{-2}} T_{m^* n}\right)\right];
\end{aligned}
\end{equation}
and \begin{equation}
\begin{aligned}
L_{s} &=\partial_{\theta}+\frac{\tilde{m}}{\sin \theta}-a \tilde{\omega} \sin \theta+s \cot \theta; \\
J_{+} &=\partial_{r}+i K / \Delta ; \quad K=\left(r^{2}+a^{2}\right)\tilde{\omega}-\tilde{m} a.
\end{aligned}
\end{equation}
Those $T_{nn}$ terms are the tetrad components of the energy momentum tensor like $T_{nn}=T_{\mu\nu}n^{\mu}n^{\nu},T_{nm^*}=T_{\mu\nu}n^{\mu}m^{\nu*}$, etc. 

Since the superradiant unstable mode is proportional to $e^{-i \omega_n t+i m \phi}$, the corresponding gravitational radiation should be monochromatic, and
\begin{equation}
    \tilde{\omega}= 2\omega,\quad \tilde{m}=\pm 2m
\end{equation}
by the property of Fourier basis. Moreover, notice that the $\mu^2$ term in Eq.~(\ref{EMTensor}) has no contribution to the Teukolsky source term because of the orthogonal relationship between null tetrads Eq.~(\ref{nulltetrads}), which could simplify the calculation of gravitational radiation.

\subsection{ Chandrasekhar-Sasaki-Nakamura Transformation and Boundary Condition}
Now we can solve the Teukolsky equation by standard Green's function method. But before that, it is worthwhile to spend some time discussing the boundary condition near the ``would-be horizon'' of a Kerr-like ECO.

Although the Teukolsky equation is powerful for the calculation of gravitational radiation, it is perhaps not the best formula to define the boundary condition of a Kerr-like ECO because it lacks a wavelike form near the ``would-be horizon''. The two linearly independent solutions of the radial Teukolsky equation near its ``would-be horizon'' are asymptotic to:
\begin{equation}
    R_{\tilde{\ell} \tilde{m} \tilde{\omega}}(r)\sim  e^{i k r^{\star}} \ \ \text{or}\ \ \Delta^{2} e^{-i k r^{\star}}
    \label{AsymptoticBehaviorPsi4}
\end{equation}
where $r^{\star}$ is the tortoise coordinate of $r$ defined by
\begin{equation}
r^{\star}=r+\frac{2 M r_{+}}{r_{+}-r_{-}} \ln \frac{r-r_{+}}{2 M}-\frac{2 M r_{-}}{r_{+}-r_{-}} \ln \frac{r-r_{-}}{2 M},
\end{equation}
and $k=\tilde{\omega}-\tilde{m} a / 2 M r_{+}$. However, we hope that the phenomenological reflectivity could intuitively show the nature of Kerr-like ECO such that it would bounce waves back from the would-be horizon. So it is necessary to define the boundary condition in the equation related to the Teukolsky equation with a standard wavelike form near the would-be horizon.
It can be realized by Chandrasekhar-Sasaki-Nakamura transformation \cite{Sasaki:2003xr,Sasaki:1981kj,Sasaki:1981sx,Chandrasekhar:1975nkd}, which could reduce to the Sasaki-Nakamura equation:
\begin{equation}
    \left[\frac{d^{2}}{d r^{* 2}}-F(r) \frac{d}{d r^{*}}-U(r)\right] X_{\tilde{\ell} \tilde{m} \tilde{\omega}}=0,
    \label{sasaki}
\end{equation}
where
\begin{equation}
\begin{aligned}
&F(r)=\frac{\eta_{,r}}{\eta} \frac{\Delta}{r^{2}+a^{2}}, \\
&\eta=c_{0}+c_{1} / r+c_{2} / r^{2}+c_{3} / r^{3}+c_{4} / r^{4}, \\
&c_{0}=-12 i \tilde{\omega} M+\tilde{\lambda}(\tilde{\lambda}+2)-12 a \tilde{\omega}(a \tilde{\omega}-\tilde{m}), \\
&c_{1}=8 i a[3 a \tilde{\omega}-\tilde{\lambda}(a \tilde{\omega}-\tilde{m})], \\
&c_{2}=-24 i a M(a \tilde{\omega}-\tilde{m})+12 a^{2}\left[1-2(a \tilde{\omega}-\tilde{m})^{2}\right] ,\\
&c_{3}=24 i a^{3}(a \tilde{\omega}-\tilde{m})-24 M a^{2} ,\\
&c_{4}=12 a^{4} .
\end{aligned}
\end{equation}
We use ``$,r$'' to abbreviate $d/dr$ here. The potential $U(r)$ is given by 
\begin{equation}
U(r)=\frac{\Delta U_{1}}{\left(r^{2}+a^{2}\right)^{2}}+G^{2}+\frac{\Delta G_{, r}}{r^{2}+a^{2}}-F G,
\end{equation}
where
\begin{equation}
\begin{aligned}
G &=-\frac{2(r-M)}{r^{2}+a^{2}}+\frac{r \Delta}{\left(r^{2}+a^{2}\right)^{2}}, \\
U_{1} &=V+\frac{\Delta^{2}}{\beta}\left[\left(2 \alpha+\frac{\beta_{, r}}{\Delta}\right)_{, r}-\frac{\eta_{, r}}{\eta}\left(\alpha+\frac{\beta_{, r}}{\Delta}\right)\right], \\
\alpha &=-i \frac{K \beta}{\Delta^{2}}+3 i K_{, r}+\lambda+\frac{6 \Delta}{r^{2}},\\
\beta &=2 \Delta\left(-i K+r-M-\frac{2 \Delta}{r}\right).
\end{aligned}
\end{equation}

With the help of this transformation, we could define the phenomenological reflectivity $\mathbf{R}$ by Sasaki-Nakamura formula:
\begin{equation}
	X_{\tilde{\ell} \tilde{m} \tilde{\omega}}^{\mathrm{ECO}} \sim e^{-i k r^*}+\mathbf{R} e^{i k r^*}\left(r \rightarrow r_{+}\right)
\end{equation}
where $0\leq|\mathbf{R}|\leq 1$. Then the corresponding homogeneous Teukolsky solution is \cite{Chen:2020htz}:
\begin{equation}
	R_{\tilde{\ell} \tilde{m} \tilde{\omega}}^{\mathrm{ECO}} \sim \frac{1}{d_{\tilde{\ell} \tilde{m} \tilde{\omega}}} \Delta^2 e^{-i k r_*}+\mathbf{R} f_{\tilde{\ell} \tilde{m} \tilde{\omega}} e^{i k r_*}\left(r \rightarrow r_{+}\right)
\end{equation}
where
\begin{equation}
\begin{aligned}
d_{\tilde{\ell} \tilde{m} \tilde{\omega}}=&\sqrt{2 r_{+}}[\left(8-24 i \tilde{\omega}-16 \tilde{\omega}^{2}\right) r_{+}^{2} \\
&+(12 i a \tilde{m}-16+16 a \tilde{m} \tilde{\omega}+24 i \tilde{\omega}) r_{+} \\
&\left.-4 a^{2} \tilde{m}^{2}-12 i a \tilde{m}+8\right], \\
f_{\tilde{\ell} \tilde{m} \tilde{\omega}}=&-\frac{4 k \sqrt{2 r_{+}}\left[2 k r_{+}+i\left(r_{+}-1\right)\right]}{\eta\left(r_{+}\right)}.
\end{aligned}
\end{equation}

\subsection{Green's Function and Energy Flux}
Now it is time to construct the Green's Function of $\phi$ of Kerr-like ECO. In the Kerr-like ECO case, we need two homogeneous solutions $\left\{R_{\tilde{\ell} \tilde{m} \tilde{\omega}}^{\text{ECO}}(r), R_{\tilde{\ell} \tilde{m} \tilde{\omega}}^{\text {up }}(r)\right\}$, which satisfy:
\begin{equation}
\begin{aligned}
&R_{\tilde{\ell} \tilde{m} \tilde{\omega}}^{\text{ECO}}(r)= \begin{cases}\dfrac{1}{d_{\tilde{\ell} \tilde{m} \tilde{\omega}}} \Delta^{2} e^{-i k r_{*}}+\mathbf{R}f_{\tilde{\ell} \tilde{m} \tilde{\omega}} e^{i k r_{*}}, & r \rightarrow r_{+}, \\ B_{\tilde{\ell} \tilde{m} \tilde{\omega}}^{\mathrm{refl}} r^{3} e^{i \tilde{\omega} r_{*}}+B_{\tilde{\ell} \tilde{m} \tilde{\omega}}^{\mathrm{inc}} r^{-1} e^{-i \tilde{\omega} r_{*}}, & r \rightarrow \infty,\end{cases}\\
&R_{\tilde{\ell} \tilde{m} \tilde{\omega}}^{\text{up}}(r)= \begin{cases}C_{\tilde{\ell} \tilde{m} \tilde{\omega}}^{\mathrm{up}} r^{3} e^{i k r_{*}}+C_{\tilde{\ell} \tilde{m} \tilde{\omega}}^{\mathrm{refl}} \Delta^{2} e^{-i k r_{*}}, & r \rightarrow r_{+}, \\ C_{\tilde{\ell} \tilde{m} \tilde{\omega}}^{\mathrm{tr} a n s} r^{3} e^{i \tilde{\omega} r_{*}}, & r \rightarrow \infty.\end{cases}
\end{aligned}
\label{HomogeneousBoundary}
\end{equation}
The corresponding Green's function is 
\begin{equation}
    \begin{aligned}
&G^{\text{ECO}}\left(r, r^{\prime}\right)=\frac{1}{W^{\text{ECO}}_{\tilde{\ell} \tilde{m} \tilde{\omega}}}  \times \\
&{\!\left[\theta\left(\!r\!-\!r^{\prime}\!\right)\!\frac{R_{\tilde{\ell} \tilde{m} \tilde{\omega}}^{\text{ECO}}\left(r^{\prime}\right)\!R_{\tilde{\ell} \tilde{m} \tilde{\omega}}^{\text{up}}(r)}{\Delta^{2}\left(r^{\prime}\right)}\!+\!\theta\!\left(\!r^{\prime}\!-\!r\!\right)\!\frac{R_{\tilde{\ell} \tilde{m} \tilde{\omega}}^{\text{ECO}}(r)\! R_{\tilde{\ell} \tilde{m} \tilde{\omega}}^{\text{up}}\left(r^{\prime}\right)}{\Delta^{2}\left(r^{\prime}\right)}\!\right]}
\end{aligned}
\label{GreensFunction}
\end{equation}
where $ W_{\tilde{\ell} \tilde{m} \tilde{\omega}}^{\text{ECO}}$ is the constant Wronskian
\begin{equation}
    W_{\tilde{\ell} \tilde{m} \tilde{\omega}}^{\text{ECO}}=2 i \tilde{\omega} C_{\tilde{\ell} \tilde{m} \tilde{\omega}}^{\text{trans}} B_{\tilde{\ell} \tilde{m} \tilde{\omega}}^{\text{inc}}.
\end{equation}
Eq.~(\ref{sourceTeukolsky}) and Eq.~(\ref{GreensFunction}) could lead to the solution of radial Teukolsky equation on the Kerr-like ECO background:
\begin{equation}
\begin{aligned}
R_{\tilde{\ell} \tilde{m} \tilde{\omega}}(r)=\dfrac{1}{W^{\text{ECO}}_{\tilde{\ell} \tilde{m} \tilde{\omega}}}\left\{ R_{\tilde{\ell} \tilde{m} \tilde{\omega}}^{\mathrm{up}} \int_{r_{+}}^{r} d r^{\prime} \frac{R_{\tilde{\ell} \tilde{m} \tilde{\omega}}^{\text{ECO}}\left(r^{\prime}\right) T_{\tilde{\ell} \tilde{m} \tilde{\omega}}\left(r^{\prime}\right)}{\Delta^{2}\left(r^{\prime}\right)} \right.\\
\left.+R_{\tilde{\ell} \tilde{m} \tilde{\omega}}^{\text{ECO}} \int_{r}^{\infty} d r^{\prime} \frac{R_{\tilde{\ell} \tilde{m} \tilde{\omega}}^{\mathrm{up}}\left(r^{\prime}\right) T_{\tilde{\ell} \tilde{m} \tilde{\omega}}\left(r^{\prime}\right)}{\Delta^{2}\left(r^{\prime}\right)}\right\}.
\end{aligned}
\end{equation}
Its  asymptotic behavior at infinity could be easily reads
\begin{equation}
\begin{aligned}
R_{\tilde{\ell} \tilde{m} \tilde{\omega}}(r \rightarrow \infty) &=\frac{r^{3} e^{i \tilde{\omega} r_{*}}}{2 i \tilde{\omega} B_{\tilde{\ell} \tilde{m} \tilde{\omega}}^{\mathrm{inc}}} \int_{r_{+}}^{\infty} d r^{\prime} \frac{T_{\tilde{\ell} \tilde{m} \tilde{\omega}}\left(r^{\prime}\right) R_{\tilde{\ell} \tilde{m} \tilde{\omega}}^{\text{ECO}}\left(r^{\prime}\right)}{\Delta^{2}\left(r^{\prime}\right)} \\
&=: Z_{\tilde{\ell} \tilde{m} \tilde{\omega}}^{\infty} r^{3} e^{i \tilde{\omega} r_{*}} .
\end{aligned}
\label{Amplitude}
\end{equation}

Since Kerr metric is asymptotic flat, $\psi_4$ is related to two independent modes of GWs, $h_+$ and $h_{\times}$, at infinity by
\begin{equation}
\psi_{4}=\frac{1}{2}\left(\ddot{h}_{+}-i \ddot{h}_{\times}\right),
\end{equation}
here a dot represents derivative of $t$. Then the energy flux of a monochromatic wave averaged over several characteristic time scale of source is given by
\begin{equation}
\left\langle\frac{d E_{GW}}{d t}\right\rangle=\sum_{\tilde{\ell}, \tilde{m}} \frac{\left|Z^{\infty}_{\tilde{\ell} \tilde{m} \tilde{\omega}}\right|^{2}}{4 \pi \tilde{\omega}^{2}}.\label{flux}
\end{equation}
Because the massive scalar source here contains only one superradiant unstable mode, and using the symmetric property of Teukolsky radial solution
\begin{equation}
    Z_{\tilde{\ell}, \tilde{m}, \tilde{\omega}}^{\infty}=Z_{\tilde{\ell},-\tilde{m},-\tilde{\omega}}^{\infty},
\end{equation}
Eq.~(\ref{flux}) can be then simplified as
\begin{equation}
\left\langle\frac{d E_{GW}}{d t}\right\rangle=2\sum_{\tilde{\ell}} \frac{\left|Z^{\infty}_{\tilde{\ell},2m,2\omega}\right|^{2}}{4 \pi (2\omega)^{2}}.
\end{equation}

\section{Numerical Results of Energy Fluxes}

\subsection{Numerical method}

The key to the  numerical calculation of the flux of an ECO-UDM system is to get the homogeneous solution on Kerr-like background with modified boundary condition. Since the reflective boundary condition is only valid for the radial part, the spin-weighted spheroidal harmonic ${}_{-2} S_{\tilde{\ell} \tilde{m}}$ and its eigenvalue is the same as the classical BH case; thus we can use Leaver's continued fractions method \cite{Leaver:1985ax,Berti:2009kk,Kokkotas:1999bd}. See Appendix A for details. For the radial part, we numerically integrate Teukolsky equation, and the modified initial condition near the ``would-be horizon'' of an ECO is given by the linear superposition of the Taylor-expanded incoming and outgoing solutions on Kerr background, namely Eq.~(\ref{ECOradial}). We set the starting point of radial integration at $r_0=r_+ +\epsilon M$, where $\epsilon=10^{-5}$.

It should be emphasized that for the scalar field on the background, we have not changed its boundary condition, i.e., it is the same as the scalar field induced by the superradiant instabilities on the BH background. We have two reasons for taking this simplified approach. Firstly, according to our previous work \cite{Guo:2021xao}, the magnitude of the eigenfrequency of UDM is not significantly affected by modifying boundary conditions for testing massive scalar field on Kerr-like ECO. Secondly, the distribution of UDM near the boundary does not significantly affect the flow of GWs corresponding to the BH case. It can be read from its radial asymptotic behavior \cite{Teukolsky:1974yv}:
\begin{equation}
    \!_{0}R_{\omega_n\ell m}(r) \sim  e^{i k r^{*}} \ \ \text{or}\ \  e^{-i k r^{*}}.
\end{equation}
Thus, unlike the case of $\psi_4$ (Eq.~(\ref{AsymptoticBehaviorPsi4})), the amplitudes of ingoing and outgoing solutions of a massive scalar field near the ``would-be horizon'' are of similar magnitudes. We conclude that the flux of the ECO-UDM system is modified mainly by corrections to the asymptotic behavior of GWs.

\subsection{Numerical Results}
In this paper we mainly focus on the dominate scalar mode in superradiant instabilities, namely the $(l,m)=(1,1$ mode.

\begin{figure}
    \centering
\subfigure[Energy fluxes of $(\tilde{l},\tilde{m})=(2,2)$ mode with $a=0$]{   
\includegraphics[width=0.5\textwidth]{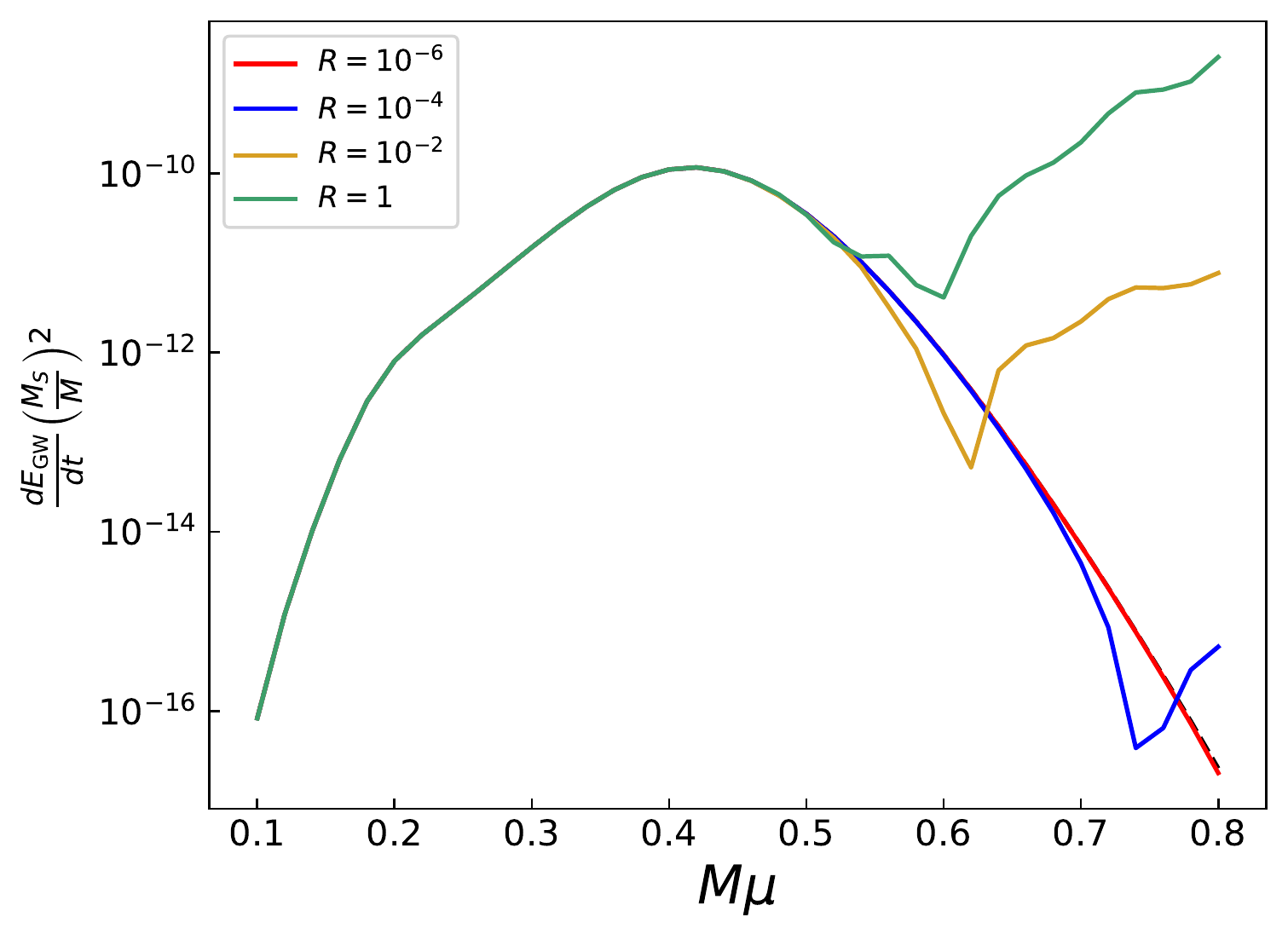}
    \label{Flux22a0}
    }
\subfigure[Energy fluxes of $(\tilde{l},\tilde{m})=(2,2)$ mode with $a=0.5$]{   
\includegraphics[width=0.5\textwidth]{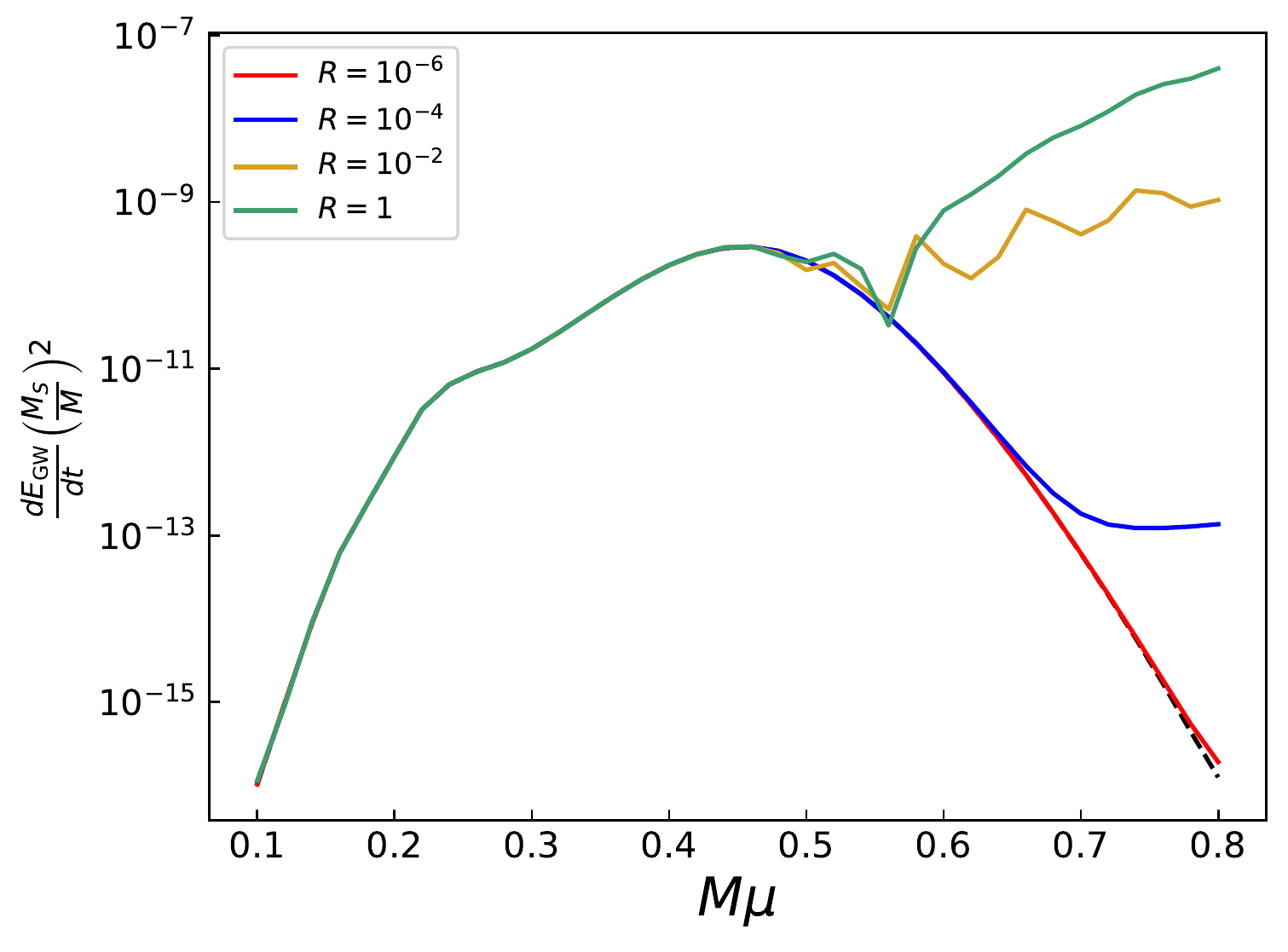}
    \label{Flux22a05}
    }
    \subfigure[Energy fluxes of $(\tilde{l},\tilde{m})=(2,2)$ mode with $a=0.9$]{   
\includegraphics[width=0.5\textwidth]{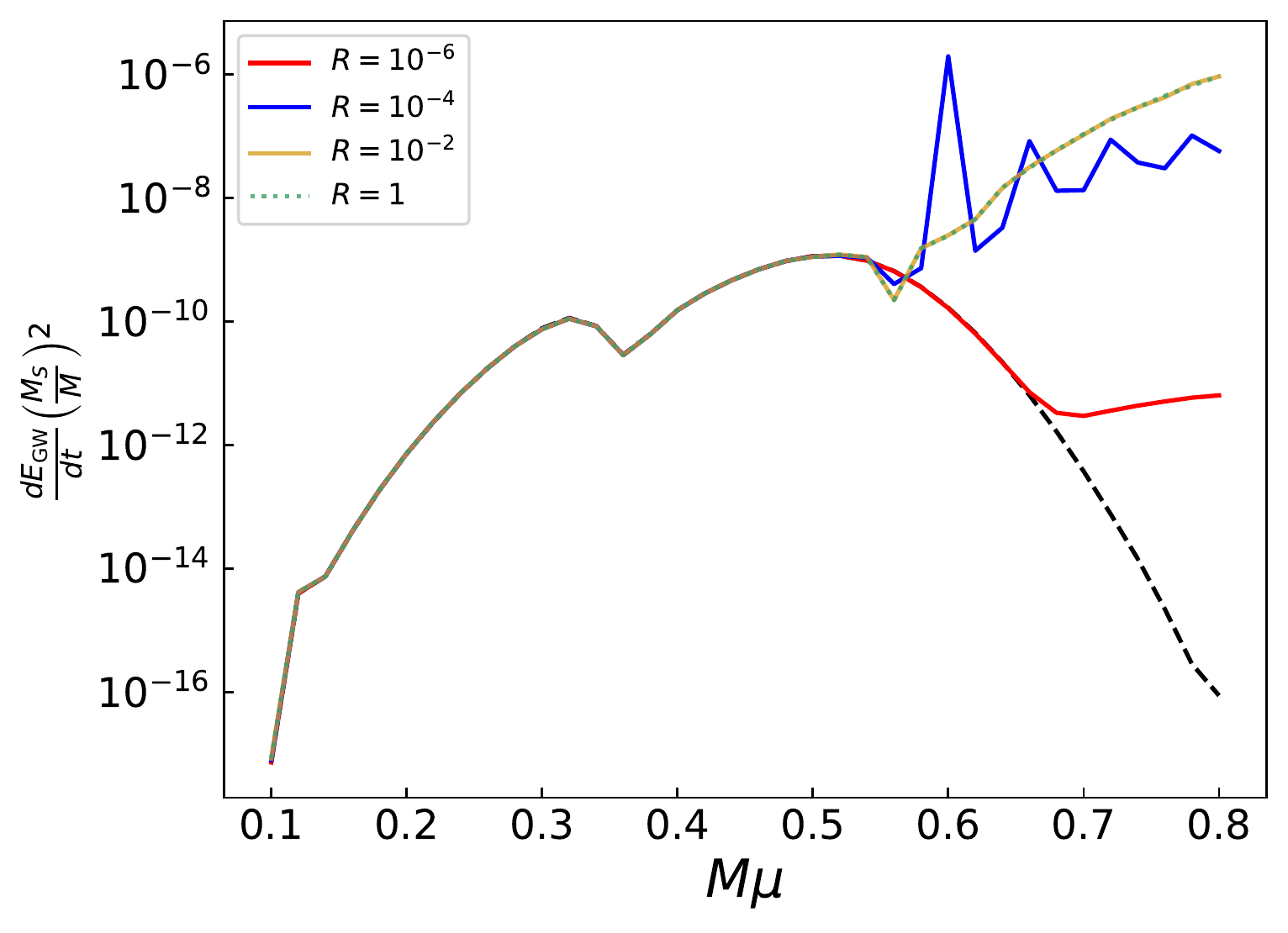}
    \label{Flux22a09}
    }
    \caption{Energy fluxes of $(\tilde{l},\tilde{m})=(2,2)$ GWs normalized by $\left(M_s / M\right)^2$ as functions of $M \mu$ with constant reflectivity.}
    \label{Flux22}
\end{figure}

In Fig.~\ref{Flux22}, we show the energy fluxes of $(\tilde{l},\tilde{m})=(2,2)$ GWs emitted by the ECO-UDM system with constant reflectivity, such that $\mathbf{R}=0$ (the BH case), $\mathbf{R}=10^{-6}$,$\mathbf{R}=10^{-4}$,$\mathbf{R}=10^{-2}$ and $\mathbf{R}=1$ (the case of full reflection) on the Kerr-like ECO background with $a=0,0.5,0.9$. Energy fluxes are normalized by $\left(M_s / M\right)^2$, where $M_s$ is the total mass of the bosonic cloud, and $M$ is the ADM
mass of the centering ECO.

We can see from Fig.~\ref{Flux22} that the energy fluxes with $M\mu\gtrsim 0.5$ (we refer to them as the large $M \mu$ cases) are sensitive to the reflectivity, while there are hardly any differences for $M\mu\lesssim0.5$ (we refer to them as the small $M \mu$ cases). This can be explained as follows. The solution for the bosonic cloud can be regarded as a hydrogen-like wave function. The larger $M\mu$ is, the closer the bosonic cloud is to the central object on average, and the more ``compact'' the bosonic cloud is. For the small $M\mu$ cases, the bosonic cloud grows mainly in the weak field region of ECO, so the physics near the event horizon has little effect on them. 

The large $M\mu$ cases are even more interesting. The general tendency of energy fluxes in the case of BHs with different $a$ is similar. And its reason has been given by \cite{Yoshino:2013ofa}. Since the bosonic clouds with large $M\mu$ are concentrated near their event horizons, the redshift effect of event horizons would become significant; hence the energy fluxes declined. While in the case of constant reflectivity, it is the presence of these reflectivities that significantly alters the physical properties near the event horizon. GWs that were supposed to go into the BH and never return will now bounce back to the observer at infinity. Thus, in the ECO case, the energy flux seen by an observer at infinity is a coherent superposition of the GWs reflected from the ``would-be horizon'' and the GWs directly propagating to infinity. It leads to the nontrivial dependence of energy flux on parameter variations.

One might be doubted by the anomaly from the monotonically increasing energy flux with reflectivity in the Fig.~\ref{Flux22}. It is because reflectivity $\mathbf{R}$ cannot be simply regarded as the energy reflectivity. It can be seen from Eq.~(\ref{HomogeneousBoundary}) that in the homogeneous case, the ratio between reflecting and ingoing energy flux should satisfy
\begin{equation}
    \dfrac{\dot{E}_{\text{reflecting}}}{\dot{E}_{\text{ingoing}}}\propto \left|f_{\tilde{\ell} \tilde{m} \tilde{\omega}}d_{\tilde{\ell} \tilde{m} \tilde{\omega}}\right|^2 \left|\mathbf{R}\right|^2
\end{equation}
up to a constant Wronskian between these two solutions. And it is consistent with the result in \cite{Xin:2021zir,Nakano:2017fvh}, in which $\mathbf{R}$ is named ``SN reflectivity'' as it is defined in Sasaki-Nakamura equations, e.g., Eq.~(\ref{sasaki}). In the inhomogeneous case, it gets more complicated. Modified physical effects near the ``would-be horizon'' of an ECO, along with the bosonic clouds concentrated nearby will produce new GWs, which exacerbates the anomaly.

\begin{figure}
    \centering
\subfigure[Energy fluxes of $(3,2)$ mode with $a=0.5$]{   
\includegraphics[width=0.5\textwidth]{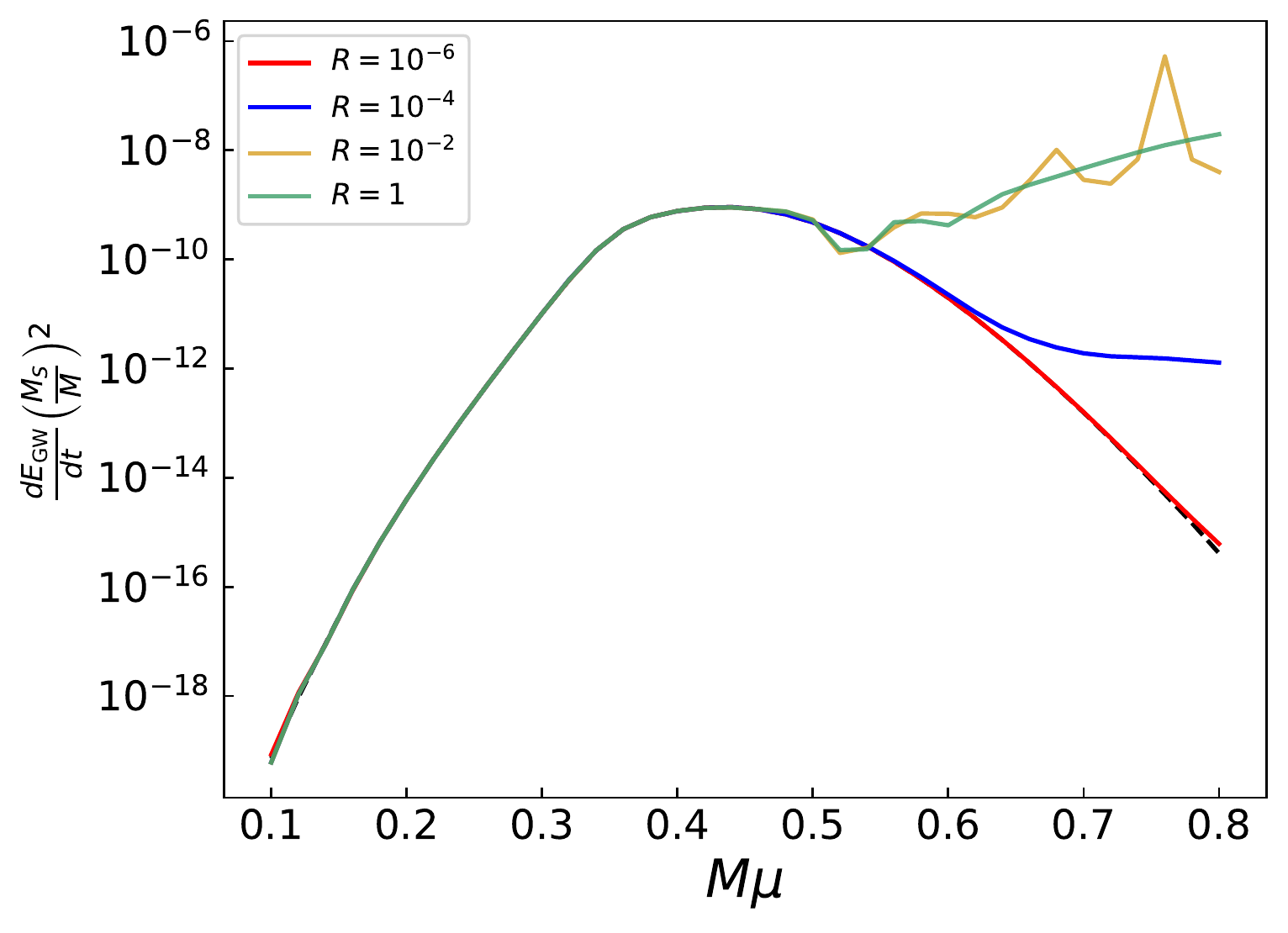}
    \label{Flux32a05}
    }
\subfigure[Energy fluxes of $(3,2)$ mode with $a=0.9$]{   
\includegraphics[width=0.5\textwidth]{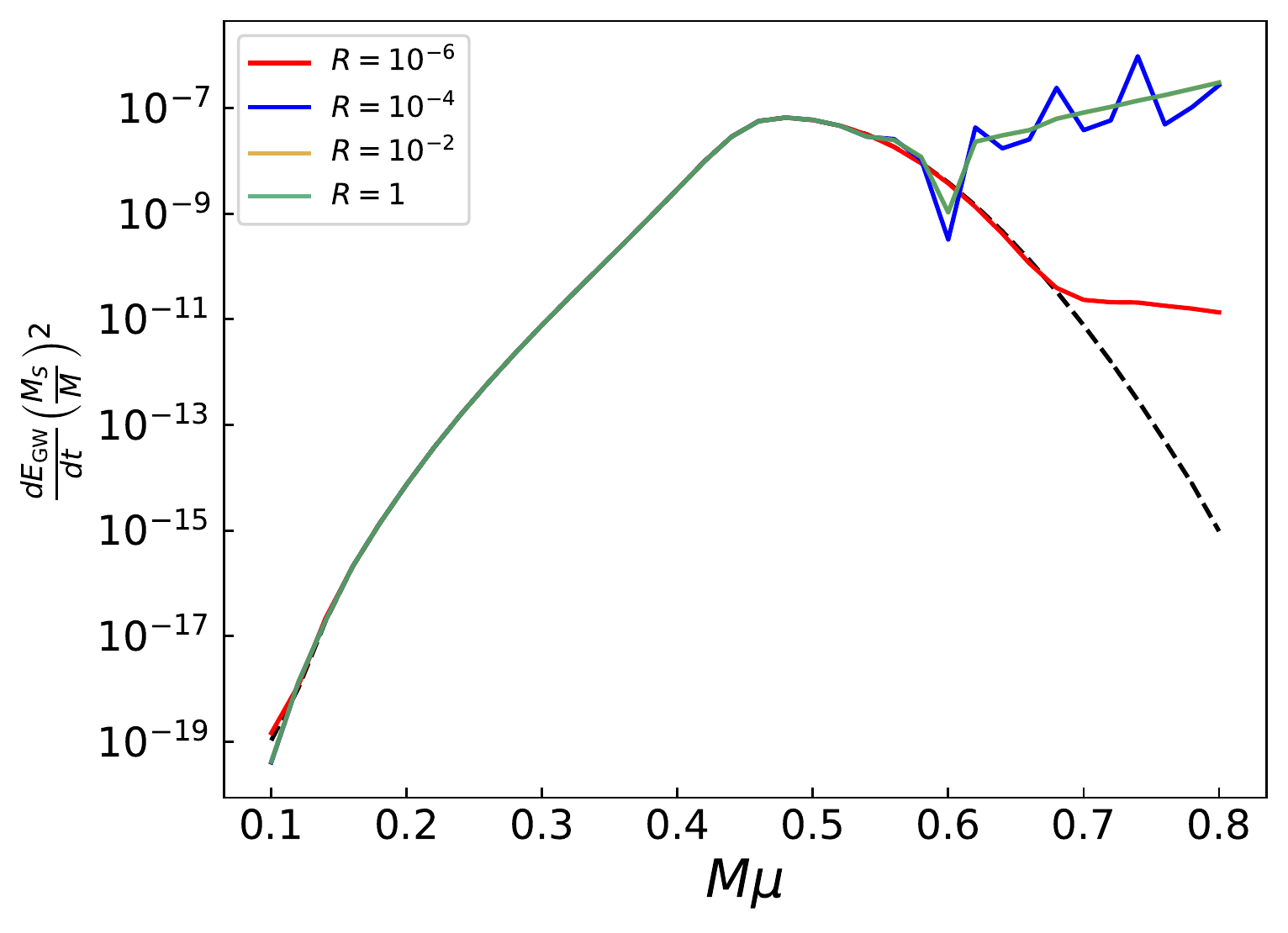}
    \label{Flux32a09}
    }
   \caption{Energy fluxes of $(3,2)$ GW mode normalized by $\left(M_S / M\right)^2$ as functions of $M \mu$ respect to different reflectivity.}
   \label{flux32}
\end{figure}

On the other hand, we plot the ecergy flux of $(\tilde{\ell},\tilde{m})=(3,2)$ mode with $a=0.5$ and $a=0.9$ in Fig.~\ref{flux32}. It can be seen from Fig.~\ref{flux32} and Fig.~\ref{Flux22} that be the behavior of energy flux with the same $\mathbf{R}$ and $a$ but with different $(\tilde{\ell},\tilde{m})$ modes can be either similar or dissimilar. The prominent examples are $\mathbf{R}=10^{-2},a=0.5$ and $\mathbf{R}=10^{-4},a=0.5$. For $\mathbf{R}=10^{-2},a=0.5$, the energy flux of GWs tend to fluctuate when $M \mu\gtrsim0.5$ for both $(\tilde{\ell},\tilde{m})=(2,2)$ mode and $(\tilde{\ell},\tilde{m})=(3,2)$ mode. However, for $\mathbf{R}=10^{-4},a=0.5$, the energy flux of GWs approaches a constant when $M \mu\gtrsim 0.5$ for the $(\tilde{\ell},\tilde{m})=(2,2)$ mode while it becomes fluctuating when $M \mu\gtrsim 0.5$ for the $(\tilde{\ell},\tilde{m})=(3,2)$ mode.


Here we should also emphasize that there is a difference between our results and the previous work \cite{Yoshino:2013ofa} when $M\mu$ becomes larger. We will discuss this problem in detail in Appendix B. In a nutshell, the calculations in \cite{Yoshino:2013ofa} are not suitable for the inhomogeneous GWs emitted by a given energy-momentum tensor on Kerr spacetime. The root of this problem has much to do with the physical pitfalls of the metric reconstruction method used in \cite{Yoshino:2013ofa}. 

The phase term $e^{i \theta}$ in $\mathbf{R}=|\mathbf{R}|e^{i \theta}$ is another valuable parameter in the reflectivity model. Its phase is often related to the location of the ``would-be horizon'' in several models. Besides, if the ECO is a wormhole, the phase term accounts for the additional delay as the waves propagate to the potential peak in the other universe and back again \cite{Mark:2017dnq}. 

\begin{figure}
    \centering
    \includegraphics[width=0.5\textwidth]{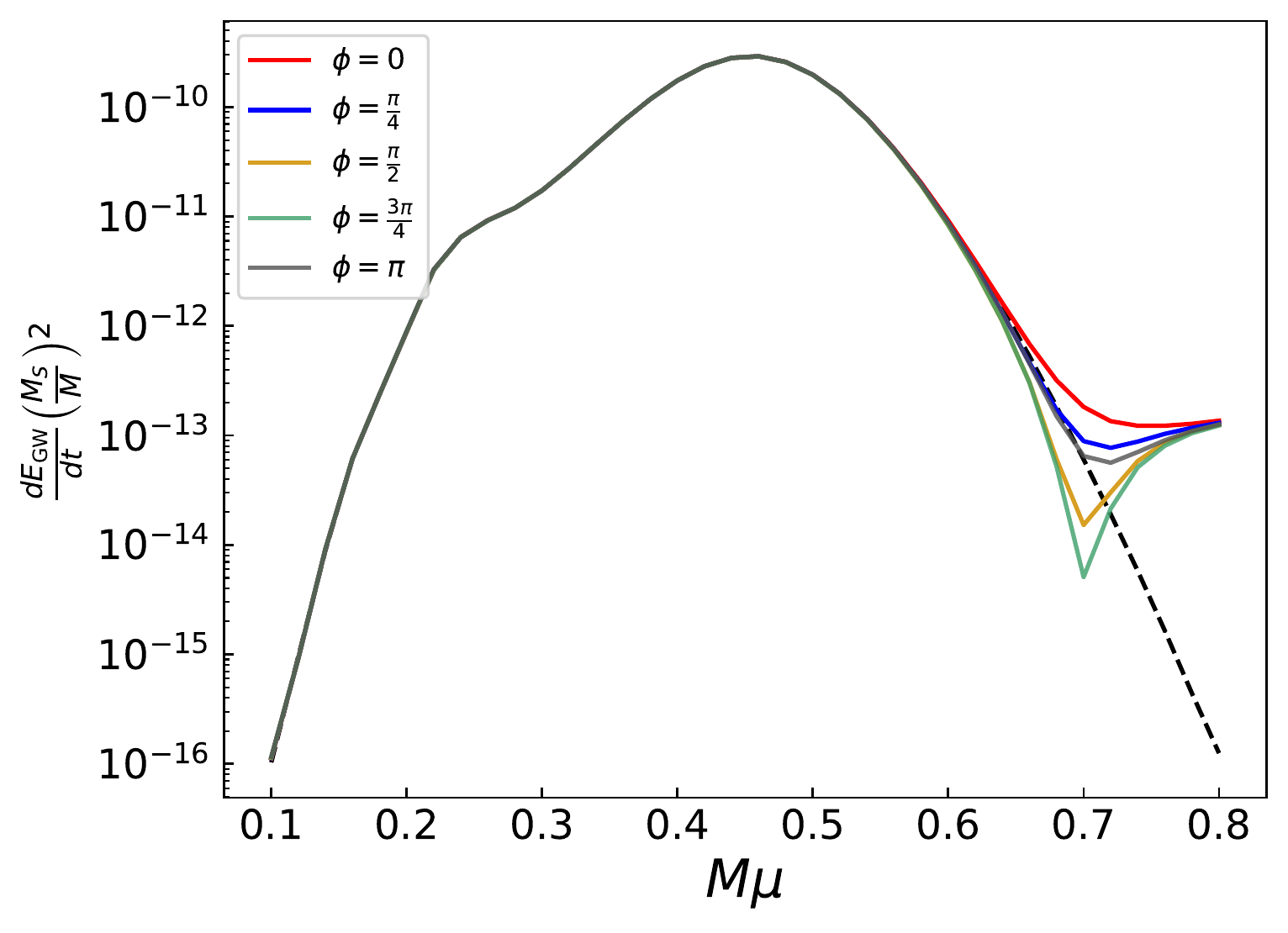}
    \caption{Energy fluxes of $(\tilde{\ell},\tilde{m})=(2,2)$ GW mode normalized by $\left(M_S / M\right)^2$ as functions of $M \mu$ respect to constant reflectivities with different phase. Here we set $a=0.5$ and $|\mathbf{R}|=10^{-4}$.}
    \label{fluxphase}
\end{figure}

\begin{figure}
    \centering
    \includegraphics[width=0.5\textwidth]{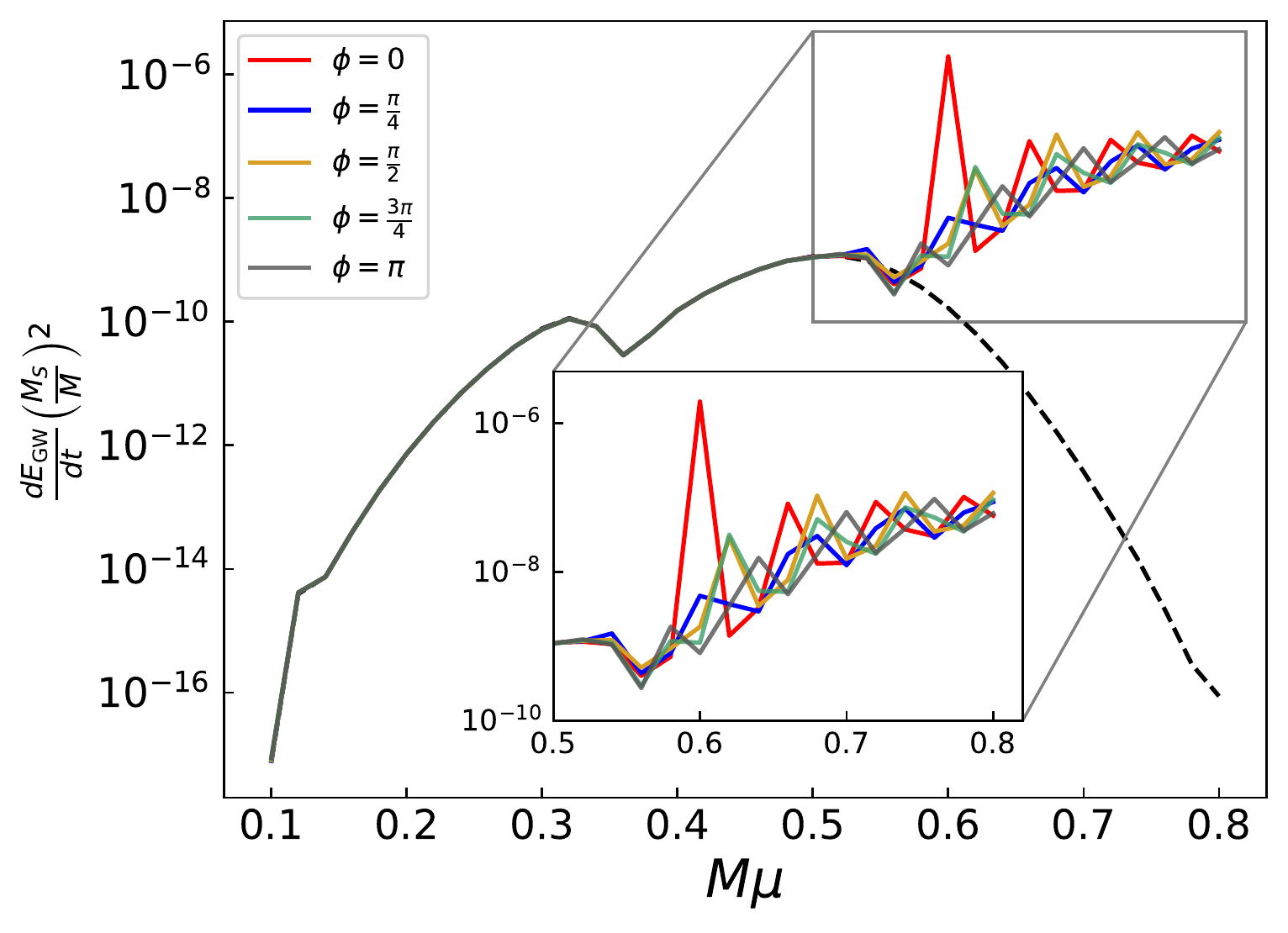}
    \caption{Energy fluxes of $(\tilde{\ell},\tilde{m})=(2,2)$ GW mode normalized by $\left(M_s / M\right)^2$  as functions of $M \mu$ respect to constant reflectivities with different phase. Here we set $a=0.9$ and $|\mathbf{R}|=10^{-4}$}
    \label{phasenew}
\end{figure}

And in the case with constant reflectivity, we can divide the previously appeared relationships into two categories: steady growth (e.g., the blue curve in Fig.~\ref{Flux22a05}) and fluctuating growth (e.g., the blue curve in Fig.~\ref{Flux22a09}). For the steady growth case, we set $(\tilde{\ell},\tilde{m})=(2,2)$, $|\mathbf{R}|=10^{-4}$ and $a=0.5$ as a benchmark example. While we choose $(\tilde{\ell},\tilde{m})=(2,2)$, $|\mathbf{R}|=10^{-4}$ and $a=0.9$ for the fluctuating growth case. The energy flux for the steady growth case with different phases is illustrated in Fig.~\ref{fluxphase}. It is not surprising that in the small $M\mu$ case, namely $M \mu \lesssim 0.5$, they share almost the same normalized energy flux. However, while $M\mu$ becomes larger, the presence of the phase term causes a well-like structure to appear in the original steady curve with $\phi=0$. With the increase of $\phi$, the corresponding energy flux will be even lower than the BH case in Fig.~\ref{Flux22a05}. 
The energy flux for the fluctuating case is shown in Fig.~\ref{phasenew}. The phase does not change the fluctuating behavior but it alters the amplitude.

Furthermore, we also consider two physical reflectivities adopted in previous studies. The first one is called ``Lorentzian reflectivity'' in \cite{Chen:2020htz}:
\begin{equation}
\mathcal{R}_{\ell m \omega}^L=\varepsilon\left(\frac{i \Gamma}{k+i \Gamma}\right) e^{-2 i b_* k},
\end{equation}
where $\varepsilon \in(0,1)$ parametrizes the amplitude reflectivity of the ECO surface, $b_*$ describes the location of the surface, and $\Gamma$ characterizes a relaxation rate of the ECO surface. It is based on the linear response theory in the framework of membrane paradigm. Since the relationship between constant reflectivity amplitude, phase term and energy flux is discussed above, we use the simplified Lorentzian reflectivity, namely $\frac{i \Gamma}{k+i \Gamma}$ below.
\begin{figure}
    \centering
    \includegraphics[width=0.5\textwidth]{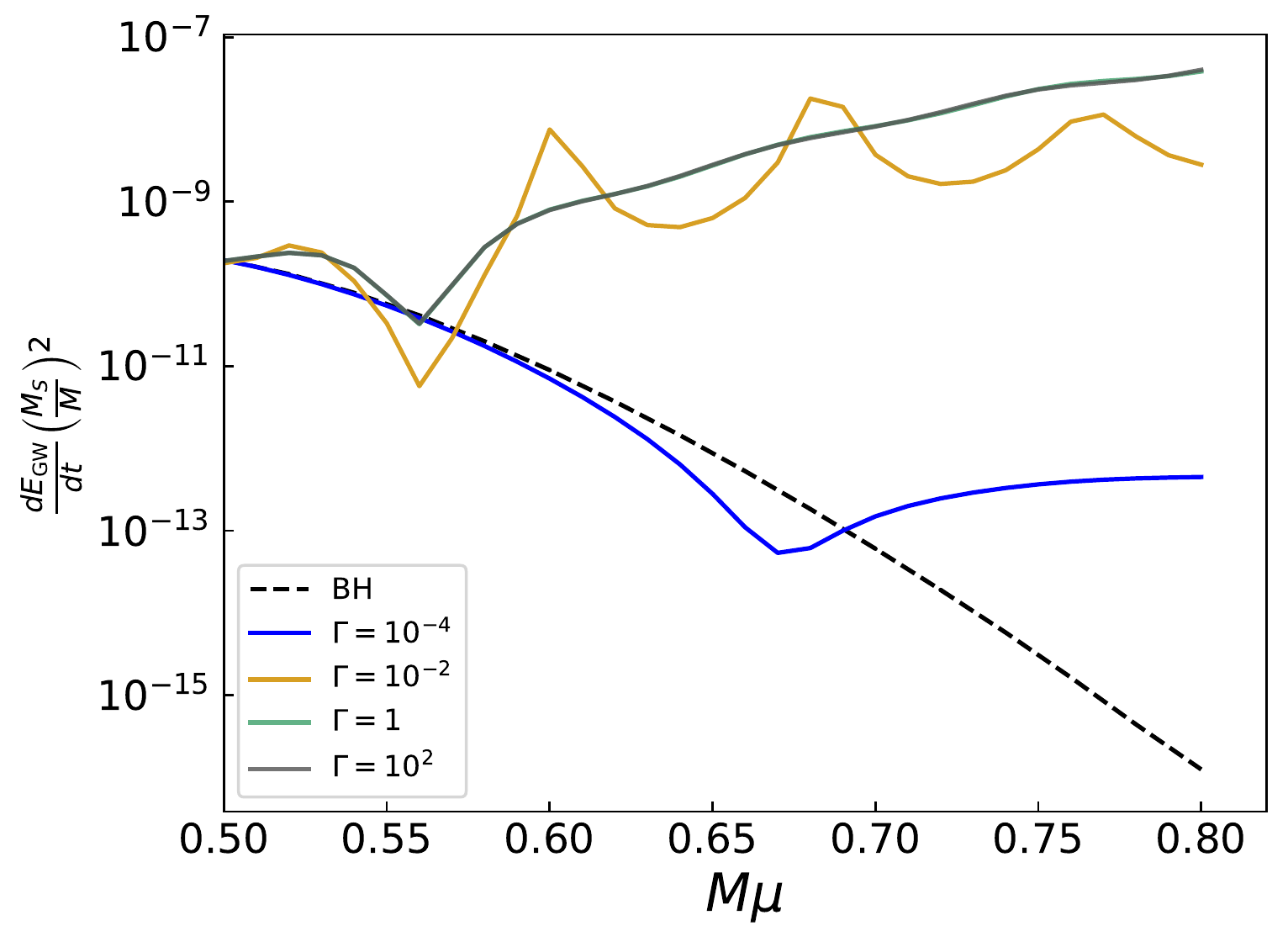}
    \caption{Energy fluxes of $(\tilde{\ell},\tilde{m})=(2,2)$ GW mode normalized by $\left(M_s / M\right)^2$ as functions of $M \mu$ respect to Lorentzian reflectivity with different relaxation coefficients $\Gamma$. Here we set $a=0.5$.}
    \label{Lorentz}
\end{figure}
In Fig.\ref{Lorentz}, we show the normalized energy fluxes with different $\Gamma$. Similar to the constant $\mathbf{R}$ model, the modification to the energy flux is only significant when $M\mu \gtrsim 0.5$. The steady and fluctuating curves also appear, and the curve returns to the case of constant $\mathbf{R}=1$ curve when  $\Gamma\gtrsim 1$.

\begin{figure}
    \centering
    \includegraphics[width=0.5\textwidth]{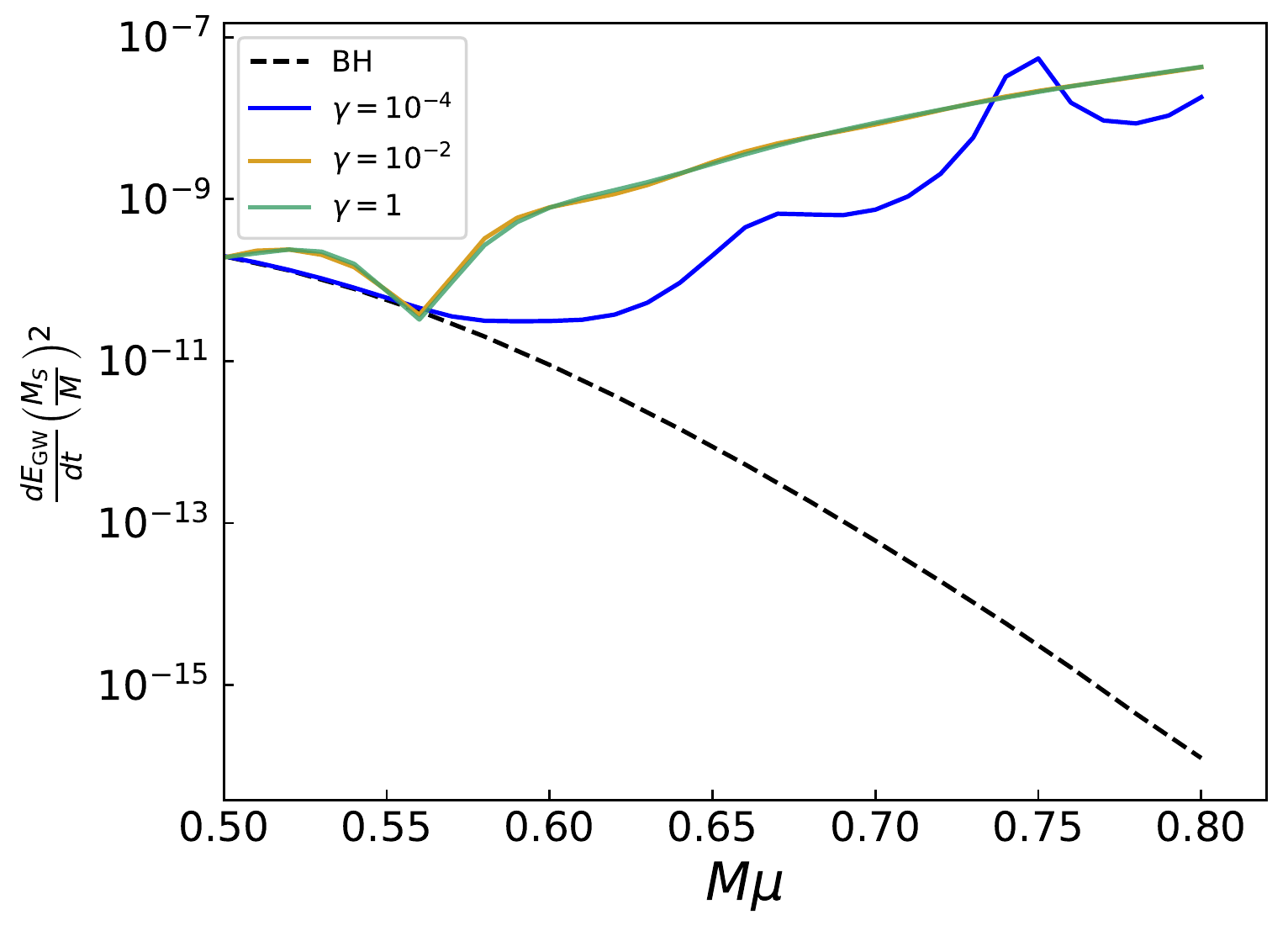}
    \caption{Energy fluxes of $(\tilde{\ell},\tilde{m})=(2,2)$ GW mode normalized by $\left(M_s / M\right)^2$ as functions of $M \mu$ respect to Boltzmann reflectivity with different free parameters $\gamma$. Here we set $a=0.5$.}
    \label{Boltzmann}
\end{figure}

The other model is named ``Boltzmann reflectivity'' \cite{Wang:2019rcf,Oshita:2019sat}:
\begin{equation}
\mathcal{R}_{\ell m \omega}^B=\exp \left(-\frac{|k|}{2 T_H}\right) \exp \left[-i \frac{|k|}{\pi T_H} \log (\gamma|k|)\right],
\end{equation}
where 
\begin{equation}
T_H=\frac{\kappa}{2 \pi}=\frac{r_{+}-r_{-}}{4 \pi\left(r_{+}^2+a^2\right)}=\frac{\sqrt{1-a^2}}{4 \pi M\left(1+\sqrt{1-a^2}\right)} 
\end{equation}
is the Hawking temperature of a Kerr BH. This reflectivity arises from the conjecture that an isolated BH should be seen as a excited multilevel quantum system; hence the detailed balance and fluctuation-dissipation theorem could be used to derive a reflectivity describing the response of this system to the external incident GWs. Here $\gamma$ is a free parameter. The results for Boltzmann reflectivity is shown in Fig.~\ref{Boltzmann}. Similar to the constant $\mathbf{R}$ model and the Lorentzian model, the modification to the energy flux is only significant when $M\mu\gtrsim 0.5$. The modified energy fluxes are several orders of magnitude larger than that of the BH case.

\section{Conclusion and Outlook}
In this paper, we calculate the energy flux of GWs from a ``UDM-ECO'' system. The spinning quantum-corrected ECO is phenomenologically described by a Kerr metric with its boundary condition changed for the test field. The energy flux is calculated by numerically solving Teukolsky equations with the source given a priori, which is a bosonic cloud around the ECO produced by the superradiant instabilities mechanism. Our results show that once the Compton wavelength of scalar UDM is close to the Schwarzschild radius of an ECO, the energy flux of this system could greatly deviate from the classical BH case. Both the amplitude and phase of the reflectivity can affect the energy flux of GWs. Besides, the relationship between energy flux and the ratio of bosonic Compton wavelength and Schwarzschild radius, reflectivity (which refers to the physical nature of an ECO), and its spin is complex, which shows the power of the coherent superposition of GWs and the new waves caused by the new physics.

Given that the bosonic cloud is usually formed far away from the central object, and the GWs produced by the bosonic cloud are dominated by $M \mu \ll 1$, hence our results may not cause strong observational effects since the modification to the GW energy flux mainly comes from $M \mu \gtrsim0.5$. However, our results show that for ECO, the new physics near its physical boundary has a significant enhancement on the overall emission of GW of the system in general. This suggests that the matter in the vicinity of the ECO has a more severe gravitational instability compared to that of a classical BH. As a result, the effect of the ECO on the accretion and radiation processes of the ultra-light dark matter field around it will be more violent. This is likely to produce deviations in the evolution of the entire scalar field above the ECO as in the BH case. We will leave it for future work.


\bigskip

{\it Acknowledgments. } 
We acknowledge the use of HPC Cluster of ITP-CAS. This work is supported by the National Key Research and Development Program of China Grant No.2020YFC2201502, grants from NSFC (grant No. 12250010,  11975019, 11991052, 12047503), Key Research Program of Frontier Sciences, CAS, Grant NO. ZDBS-LY-7009, CAS Project for Young Scientists in Basic Research YSBR-006, the Key Research Program of the Chinese Academy of Sciences (Grant NO. XDPB15).

	
\bibliographystyle{apj}
\bibliographystyle{apsrev4-1}
\bibliography{DMReflect}
\begin{appendices}
\section{Leaver’s continued fractions method}
In this appendix, we briefly introduce Leaver’s continued fractions
method to construct the UDM cloud and spin-weighted spheroidal harmonics on Kerr background \cite{Leaver:1985ax, Berti:2005gp}. Both the Teukolsky radial and the angular equation can be transformed into the standard confluent Heun equation, which has two regular singular points and one irregular singular point. And the calculation of its eigenvalue and eigenfunction with suitable boundary conditions can be reduced to a central two-point connection problem, which can be solved using the well-developed technique of series expansions by mathematicians.

For the Teukolsky angular equation 
\begin{equation}
\left[\left(1-u^2\right) S_{l m, u}\right]_{, u}+\left[a^2 \omega^2 u^2-2 a \omega s u +s+ \ _sA_{l m}-\frac{(m+s u)^2}{1-u^2}\right] S_{l m}=0,
\end{equation}
where $u=\cos \theta$, its eigenfunction can be expanded as
\begin{equation}
S_{l m}(u)=\mathrm{e}^{a \omega u}(1+u)^{\frac{1}{2}|m-s|}(1-u)^{\frac{1}{2}|m+s|} \sum_{n=0}^{\infty} a_n(1+u)^n
\end{equation}

Since the series appearing above should be convergent, we can get the recurrence relation
\begin{equation}
\left.\begin{array}{c}
\alpha_0^\theta a_1+\beta_0^\theta a_0=0 \\
\alpha_n^\theta a_{n+1}+\beta_n^\theta a_n+\gamma_n^\theta a_{n-1}=0 \quad n=1,2 \ldots
\end{array}\right\},
\end{equation}
and the recurrence coefficients are
\begin{equation}
\begin{aligned}
\alpha_n^\theta=&-2(n+1)\left(n+2 k_1+1\right) \\
\beta_n^\theta=& n(n-1)+2 n\left(k_1+k_2+1-2 a \omega\right) \\
&-\left[2 a \omega\left(2 k_1+s+1\right)-\left(k_1+k_2\right)\left(k_1+k_2+1\right)\right]\\
&-\left[a^2 \omega^2+s(s+1)+\ _sA_{l m}\right] \\
\gamma_n^\theta=& 2 a \omega\left(n+k_1+k_2+s\right) .
\end{aligned}
\end{equation}
where
\begin{equation}
k_1=\frac{1}{2}|m-s| , k_2=\frac{1}{2}|m+s| .
\end{equation}

The eigenvalue $\ _sA_{l m}$ is determined by the recurrence relation, and it can be reduced to the continued fraction equation
\begin{equation}
0=\beta_0^\theta-\frac{\alpha_0^\theta \gamma_1^\theta}{\beta_1^\theta-} \frac{\alpha_1^\theta \gamma_2^\theta}{\beta_2^\theta-} \frac{\alpha_2^\theta \gamma_3^\theta}{\beta_3^\theta-} \ldots
\end{equation}
where
\begin{equation}
\frac{\alpha_0^\theta \gamma_1^\theta}{\beta_1^\theta-} \frac{\alpha_1^\theta \gamma_2^\theta}{\beta_2^\theta-} \frac{\alpha_2^\theta \gamma_3^\theta}{\beta_3^\theta-} \ldots=\dfrac{\alpha_0^\theta \gamma_1^\theta}{\beta_1^\theta-\dfrac{\alpha_1^\theta \gamma_2^\theta}{\beta_2^\theta-\dfrac{\alpha_2^\theta \gamma_3^\theta}{\beta_3^\theta-\ldots}}}.
\end{equation}

This method can be effective when calculating $\psi_4$ because the frequency $\omega_{GW}$ is fixed before this procedure. However, in the scalar case, both $\omega$ and $\ _sA_{l m}$ should be determined simultaneously. Thus it is better to support a explicit formula related to $\omega$.

Here we use the small-$a\omega$ expansion of $\ _sA_{l m}$ \cite{Berti:2005gp}:
\begin{equation}
{ }_s A_{l m}=\sum_{p=0}^{\infty} f_p (a\omega)^p.
\end{equation}
By introducing
\begin{equation}
\begin{aligned}
&\frac{1}{2}(\alpha+\beta)=\max (|m|,|s|),\\ &\frac{1}{2}(\alpha-\beta)=\frac{m s}{\max (|m|,|s|)}, \\
&h(l)=\frac{\left[l^2-\frac{1}{4}(\alpha+\beta)^2\right]\left[l^2-\frac{1}{4}(\alpha-\beta)^2\right]\left(l^2-s^2\right)}{2\left(l-\frac{1}{2}\right) l^3\left(l+\frac{1}{2}\right)}
\end{aligned},
\end{equation}
we have 
\begin{equation}
\begin{aligned}
f_0=&l(l+1)-s(s+1)\\
f_1=&-\frac{2 m s^2}{l(l+1)}\\
f_2=&h(l+1)-h(l)-1\\
f_3=&\frac{2 h(l) m s^2}{(l-1) l^2(l+1)}-\frac{2 h(l+1) m s^2}{l(l+1)^2(l+2)}\\
f_4=& m^2 s^4\left(\frac{4 h(l+1)}{l^2(l+1)^4(l+2)^2}-\frac{4 h(l)}{(l-1)^2 l^4(l+1)^2}\right)\\
& -\frac{(l+2) h(l+1) h(l+2)}{2(l+1)(2 l+3)}+\frac{h^2(l+1)}{2(l+1)}\\
& +\frac{h(l) h(l+1)}{2 l^2+2 l}-\frac{h^2(l)}{2 l}+\frac{(l-1) h(l-1) h(l)}{4 l^2-2 l}\\
f_5=& m^3 s^6\left(\frac{8 h(l)}{l^6(l+1)^3(l-1)^3}-\frac{8 h(l+1)}{l^3(l+1)^6(l+2)^3}\right)\\
& +m s^2 h(l)\left(-\frac{h(l+1)\left(7 l^2+7 l+4\right)}{l^3(l+2)(l+1)^3(l-1)}\right)\\
& +m s^2 h(l)\left(-\frac{h(l-1)(3 l-4)}{l^3(l+1)(2 l-1)(l-2)}\right)\\
&+ m s^2\left(\frac{(3 l+7) h(l+1) h(l+2)}{l(l+1)^3(l+3)(2 l+3)}\right.\\
& \left.-\frac{3 h^2(l+1)}{l(l+1)^3(l+2)}+\frac{3 h^2(l)}{l^3(l-1)(l+1)}\right)\\
\ldots
\end{aligned}.
\end{equation}

For the Teukolsky radial equation for massive scalar field (Eq.~\ref{scalareq}), Dolan \cite{Dolan:2007mj} has proved that the radial eigenfunction can be expressed as
\begin{equation}
R(r)=\left(r-r_{+}\right)^{-i \sigma}\left(r-r_{-}\right)^{i \sigma+\chi-1} e^{-k r} \sum_{n=0}^{\infty} d_n\left(\frac{r-r_{+}}{r-r_{-}}\right)^n
\end{equation}
where
\begin{equation}
\sigma=\frac{2 M r_{+}}{r_{+}-r_{-}}\left(\omega-m \Omega_H\right), \quad \chi=\frac{M\left(2 \omega^2-\mu^2\right)}{k}.
\end{equation}

The convergence condition of this series leads to the three-term
recurrence relation
\begin{equation}
\begin{gathered}
d_1=-\frac{\beta_0^r}{\alpha_0^r} d_0, \\
\alpha_n^r d_{n+1}+\beta_n^r d_n+\gamma_n^r d_{n-1}=0,
\end{gathered}
\end{equation}
where
\begin{equation}
\begin{aligned}
&\alpha_n=n^2+\left(c_0+1\right) n+c_0, \\
&\beta_n=-2 n^2+\left(c_1+2\right) n+c_3, \\
&\gamma_n=n^2+\left(c_2-3\right) n+c_4 .
\end{aligned}
\end{equation}
with
\begin{equation}
\begin{aligned}
c_0=& 1-2 i \omega-\frac{2 i}{b}\left(\omega-\frac{a m}{2}\right), \\
c_1=&-4+4 i(\omega-i q(1+b))+\frac{4 i}{b}\left(\omega-\frac{a m}{2}\right)-\frac{2\left(\omega^2+q^2\right)}{q}, \\
c_2=& 3-2 i \omega-\frac{2\left(q^2-\omega^2\right)}{q}-\frac{2 i}{b}\left(\omega-\frac{a m}{2}\right) \\
c_3=& \frac{2 i(\omega-i q)^3}{q}+2(\omega-i q)^2 b+q^2 a^2+2 i q a m-A_{l m}-1\\
&-\frac{(\omega-i q)^2}{q}+2 q b +\frac{2 i}{b}\left(\frac{(\omega-i q)^2}{q}+1\right)\left(\omega-\frac{a m}{2}\right), \\
c_4=& \frac{(\omega-i q)^4}{q^2}+\frac{2 i \omega(\omega-i q)^2}{q}-\frac{2 i}{b} \frac{(\omega-i q)^2}{q}\left(\omega-\frac{a m}{2}\right) .
\end{aligned}
\end{equation}
and $b=\sqrt{1-a^2}$.

The frequency of the massive scalar field can be induced from the continued fraction equation
\begin{equation}
\beta_0^r-\frac{\alpha_0^r \gamma_1^r}{\beta_1^r-} \frac{\alpha_1^r \gamma_2^r}{\beta_2^r-} \frac{\alpha_2^r \gamma_3^r}{\beta_3^r-} \ldots=0.
\end{equation}

\section{Metric Reconstruction and Energy Fluxes}
The calculation of energy flux produced by a given source on  kerr background in \cite{Yoshino:2013ofa} is heavily based on the metric reconstruction procedure \cite{Chrzanowski:1975wv,Kegeles:1979an}, which is also called CCK procedure. Particularly, in \cite{Chrzanowski:1975wv}, Chrzanowski showed that the homogeneous metric perturbation $h_{\mu\nu}$ on a BH can be determinded by acting a linear differential operator, named Chrzanowski operator below,  to the homogeneous solutions for Newman-Penrose scalars $\psi_0$ and $\psi_4$. And Wald \cite{Wald:1978vm} proved it in a much more concise and elegant way.

Although the mathematical correctness of homogeneous metric reconstruction is well documented, it is warned in \cite{Wald:1978vm} that there exists the ``physical risks'' arising from this procedure. In \cite{Ori:2002uv}, Ori emphasized that while using $\psi_0$ or $\psi_4$ to construct $h_{\mu\nu}$, they may not share the same physical meaning.

It is helpful to revisit the process of CCK procedure. We introduce these four partial linear differential operators:
\begin{equation}
\begin{aligned}
    \mathcal{E}\quad & \text{Einstein operator. The equation of metric}\\ &\text{perturbation can be pressed as}\  \mathcal{E}h_{\mu\nu}=0.\\
    \mathcal{O}\quad & \text{Teukolsky operator for }\psi_0\text{. The equation of}\\
    &\text{Teukolsky equation can be pressed as}\  \mathcal{O}\psi_0=0.\\
    \mathcal{T}\quad & \text{"Extracting }\psi_0\text{" operator. A given  metric}\\ &\text{perturbation can be transformed into its}\\
    &\text{corresponding }\psi_0\text{ as }\mathcal{T}h_{\mu\nu}=\psi_0.\\
    \mathcal{S}\quad & \text{Adjoint Chrzanowski operaor, which will be}\\
    &\text{discussed bellow. It satisfies }\mathcal{S}\mathcal{E}=\mathcal{O}\mathcal{T}
\end{aligned}
\end{equation}

Now consider their adjoint operator, which is defined as for any $m-$index tensor field $\phi_{\mu_1\mu_2\cdots\mu_m}$ and $\xi_{\mu_1\mu_2\cdots\mu_m}$, the adjoint linear differential operator $\mathcal{V}^{\dagger}$ of $\mathcal{V}$ should satisfy
\begin{equation}
    \xi^{\mu_1\mu_2\cdots\mu_m}\mathcal{V}\phi_{\mu_1\mu_2\cdots\mu_m}-\left(\mathcal{V}^{\dagger} \xi^{\mu_1\mu_2\cdots\mu_m}\right)\phi_{\mu_1\mu_2\cdots\mu_m}=\nabla_{\mu}t^{\mu}.
\end{equation}
One can check that $ \mathcal{E}^{\dagger}=\mathcal{E}$, e.g., it is self-adjoint operator. And $\mathcal{O}^{\dagger}$ is just the Teukolsky operator for $\rho^{-4} \psi_4$. Wald have proved such mathematical theorem \cite{Wald:1978vm}: Suppose the identity $\mathcal{S}\mathcal{E}=\mathcal{O}\mathcal{T}$ holds for some linear partial differential operators $\mathcal{S},\mathcal{E}$, $\mathcal{O}$, and $\mathcal{T}$. Suppose some $\psi$ is the solution of $\mathcal{O}^{\dagger} \psi=0$. Then $\mathcal{S}^{\dagger} \psi$ satisfies $\mathcal{E}^{\dagger}\left(\mathcal{S}^{\dagger} \psi\right)=0$. Thus, in particular, if $\mathcal{E}$ is self-adjoint then $\mathcal{S}^{\dagger} \psi$ is a solution of $\mathcal{E}(f)=0$.Using this theorem, one can easily find that given a solution $\Psi$ of Teukolsky equation for $\rho^{-4}\psi_4$, $\mathcal{S}^{\dagger}\Psi$ is the vacuum metric linear perturbation on Kerr background. We call $\mathcal{S}^{\dagger}$ as Chrzanowski operator. And this becomes a metric reconstruction procedure.

However, we cannot extract the same $\rho^{-4}\psi_4$ as $\Psi$! In particular, suppose $\mathcal{T}'$ is the  "extracting $\rho^{-4}\psi_4$ operator", we cannot derive out
\begin{equation}
    \mathcal{T}'\mathcal{S}^{\dagger}\Psi=\Psi.
\end{equation}
It means that $\Psi$ and its corresponding metric perturbation have different physical meaning. Since the calculation of energy flux produced by a given source on Kerr background needs to integrate metric perturbation outside the event horizon, this procedure would lead to different energy flux compared to the result of Teukolsky method.

And it will be clearer to directly compare the calculation results under both formulas in the BH case. The reconstructed metric outside a bounded source is given by
\begin{equation}
h_{\mu \nu}=\int_{-\infty}^{\infty} d \omega \sum_{i, m} \sum_P \frac{2 i \omega}{|\omega|} h_{\mu \nu}^{\mathrm{up}}(l m \omega P)\left\langle h_{\alpha \beta}^{\text {out }}(l m \omega P), T^{\alpha \beta}\right\rangle
\end{equation}
and the corresponding energy flux calculated by CCK procedure is
\begin{equation}
    \frac{dE_{GW}}{dt}=\sum_{lmP}\int_0^{\omega}|\left<h_{\alpha\beta}^{out}(lm\omega P),T^{\alpha\beta}\right>|^2\omega d\omega,
    \label{CCKflux}
\end{equation}
where
\begin{equation}
    \left<h_{\alpha\beta},T^{\alpha\beta}\right>=\int h^{*}_{\alpha\beta}T^{\alpha\beta} \sqrt{-g}d^4 x
\end{equation}
is the inner product of $h_{\alpha\beta}$, $T^{\alpha\beta}$. And $P=\pm 1$ denotes the parity of GW.

Using the CCK procedure, one can construct $h_{\alpha\beta}^{out}(lm\omega P)$ by
\begin{equation}
\begin{aligned}
h^{out}_{\mu \nu}(lm\omega P) &=\hat{U}_{+2} R_{lm}^{\omega\ \text{out}}(r)_{-2} S_{l m}^{\omega}(\theta) e^{i m \phi-i \omega t} \\
&+P\hat{U}^{*} _{+2} R_{lm}^{\omega\ \text{out}}(r)_{+2} S_{l m}^{\omega}(\theta) e^{i m \phi-i \omega t}
\end{aligned}
\end{equation}
in the outgoing gauge $h_{\mu \nu} n^\nu=h_\nu^\nu=0$, where
\begin{equation}
\begin{aligned}
\hat{U}=&-n_\mu n_\nu \mathcal{A}-m_\mu m_\nu \mathcal{B}+n_{(\mu} m_{\nu)} \mathcal{C}\\
\mathcal{A}=& \rho^{*-4}\left(\delta-3 \alpha^*-\beta+5 \pi^*\right)\left(\delta-4 \alpha^*+\pi^*\right) \\
\mathcal{B}=& \rho^{*-4}\left(\Delta+5 \mu^*-3 \gamma^*+\gamma\right)\left(\Delta+\mu^*-4 \gamma^*\right) \\
\mathcal{C}=& \rho^{*-4}\left[\left(\delta+5 \pi^*+\beta-3 \alpha^*+\tau\right)\left(\Delta+\mu^*-4 \gamma^*\right)\right.\\
&\left.\quad+\left(\Delta+5 \mu^*-\mu-3 \gamma^*-\gamma\right)\left(\delta-4 \alpha^*+\pi^*\right)\right] .
\end{aligned}
\end{equation}
Here, $\alpha, \beta, \gamma, \mu$, and $\pi$ are the Newman-Penrose variables (see \cite{teukolsky_1972,Chandrasekhar:1985kt}), and $\Delta=n^\mu \nabla_\mu, \delta=m^\mu \nabla_\mu$ are the directional derivatives along with null tetrads. From now on, we use $\Delta_K$ to denote $r^2-2M r+a^2$ to avoid unnecessary confusion.

For the Teukolsky formula, according to Eq.~(\ref{TRE}),Eq.~(\ref{flux}) and Eq.~(\ref{Amplitude}), we get
\begin{equation}
	\frac{d E_{G W}}{d t}=\sum_{l m} \int d \omega \frac{1}{16 \pi \omega^4} \frac{\left|\int_{r_{+}}^{\infty}{ }_{-2} R_{l m}^{\omega \text { in }} \Delta_K^{-2}{ }_{-2} T_{l m \omega}(r) d r\right|^2}{\left|d_{l m \omega} B_{l m}^{i n c}\right|^2}
\end{equation}
where 
${ }_{-2} R_{l m}^\omega$ in and $d_{l m \omega} B_{l m}^{i n c}$ correspond to our $R_{\tilde{\ell} \tilde{m} \tilde{\omega}}^{\mathrm{ECO}}(r)$ respect to zero reflectivity. Here we use the Newman-Penrose variables to represent the source term:
\begin{equation}
\begin{aligned}
\frac{\rho^{4}}{2}{}_{-2} T=& \left(\Delta+3 \gamma-\gamma^*+4 \mu+\mu^*\right)\left(\delta^*-2 \tau^*+2 \alpha\right) T_{n m^*}\\
-&\left(\Delta+3 \gamma-\gamma^*+4 \mu+\mu^*\right)\left(\Delta+2 \gamma-2 \gamma^*+\mu ^{*}\right) T_{m^* m^*}\\
+&\left(\delta^*-\tau^*+\beta^*+3 \alpha+4 \pi\right)\left(\Delta+2 \gamma+2 \mu^*\right) T_{n m^*}\\
-&\left(\delta^*-\tau^*+\beta^*+3 \alpha+4 \pi\right)\left(\delta^*-\tau^*+2 \beta^*+2 \alpha\right)T_{nn}.
\end{aligned}
\end{equation}
and
\begin{equation}
    \ _{-2}T_{lm\omega}(r)=2\int d\Omega dt \Sigma_K {}_{-2}S^{\omega}_{lm}(\theta)e^{i\omega t-im\phi} {}_{-2}T
\end{equation}
Comparing these two formula of energy flux, it can be found that it is the inner product term in Eq.(\ref{CCKflux}) that determines how the homogeneous solution interact with source term. Therefore, we focus on this inner product bellow.

Notice that these linear operators act on different functions, it is convient to treat them in a samilar manner, e.g., operators acting on energy-momentum term. And we have to do integration by parts, which is to compute the adjoint operators of these linear operators. Since they all have the structure
\begin{equation}
    \mathcal{L}=\xi^{\mu}\nabla_{\mu}+c,
\end{equation}
where $c$ is some scalar. Their adjoint operators should be
\begin{equation}
    \mathcal{L}^{\dagger}=-\mathcal{L}+c-\nabla_{\mu}\xi^\mu.
\end{equation}

So, by using Newman-Penrose transportation equation \cite{Chandrasekhar:1985kt}, we can get
\begin{equation}
    \begin{aligned}
    \Delta+c &\stackrel {Adjoint}{\longrightarrow} -\Delta+c-\left(\gamma+
    \gamma^*+\mu+\mu^*\right)\\
    \delta+c &\stackrel {Adjoint}{\longrightarrow} -\delta+c-\left(\pi^*+\tau+\beta-\alpha^*\right)\\
    \delta^*+c &\stackrel {Adjoint}{\longrightarrow} -\delta^*+c-\left(\pi+\tau^*+\beta^*-\alpha\right).\\
    \end{aligned}
\end{equation}

The other thing to deal with is the spin-weighted functions. We can use the symmetric properties of Teukolsky solutions:
\begin{equation}
\begin{aligned}
&{ }_s R_{l m }^{\omega}(r)=(-)^m{ }_s R^{-\omega}_{l-m}(r)^* \\
&{ }_s R^{\omega}_{l m }(r)^*=\left(1 / \Delta_K\right)^s{ }_{-s} R^{\omega}_{l m }(r),\\
&{}_s S^{\omega}_{lm}(\theta)=(-1)^m {}_{-s}S_{l-m}^{-\omega}(\theta)
\end{aligned}
\end{equation}
one can get
\begin{equation}
\begin{aligned}
    &{}_{+2}R^{\omega\ \text{out}}_{lm}(r)^{*}=\Delta_K^{-2}{}_{-2}R^{\omega\ \text{in}}_{lm}(r)\\
    &{}_{+2}S^{\omega}_{lm}(\theta)e^{im\phi-i\omega t}=(-1)^m \left({}_{-2}S^{\omega}_{lm}(\theta)e^{im\phi-i\omega t}\right)^* .
\end{aligned}
\end{equation}
In the end, we can get the inner product term:
\begin{equation}
\begin{aligned}
    \left<h_{\alpha\beta}^{out}(lm\omega P),T^{\alpha\beta}\right>=&\int_{r_+}^{\infty}\Delta_K^{-2}{}_{-2}R^{\omega\ \text{in}}_{lm}(r)\tilde{T}\\
    +&P (-1)^m\int_{r_+}^{\infty}\Delta_K^{-2}{}_{-2}R^{\omega\ \text{in}}_{lm}(r)\tilde{T}^{*},
    \end{aligned}
\end{equation}
where
\begin{equation}
    \tilde{T}^{*}=\int d\Omega dt\Sigma_K {}_{-2}S^{\omega}_{lm}(\theta)e^{i\omega t-im\phi} \tilde{\tilde{T}}
\end{equation}
and
\begin{equation}
\begin{aligned}
    \tilde{\tilde{T}}=&\left(\delta^*+3\alpha+\tau^*+\beta^*\right)\left(\Delta-4\mu+2\mu^*+4\gamma+2\gamma^*\right)\rho^{-4}T_{nm^*}\\
    -&\left(\delta^*+3\alpha+\tau^*+\beta^*\right)\left(\delta^*+2\alpha+2\beta^*-4\pi+\tau^*\right)\rho^{-4}T_{nn}\\
    +&\left(\Delta+5\gamma+\gamma^*+\mu^*\right)\left(\delta^*-4\pi+2\alpha\right)\rho^{-4}T_{nm^*}\\
    -&\left(\Delta+5\gamma+\gamma^*+\mu^*\right)\left(\Delta-4\mu+4\gamma-\mu^*\right)\rho^{-4}T_{m^*m^*}.
    \end{aligned}
\end{equation}

Although these two formula is similar, one can easily check that there are key differences in their source terms. This is consistent with our previous argument that traditional CCK procedure may lead to diffrent physical case.

For the above reasons, and the correctness of Teukolsky method has been widely demonstrated, we use the calculation directly based on $\rho^{-4}\psi_4$ in the main text instead of the method in \cite{Yoshino:2013ofa}, although this result can be used to get the analytical flat approximation and  to discuss the contribution of different parity to energy flow. The energy flux calculated by metric reconstruction method may better use the promoted CCK procedure for inhomogeneous case \cite{Ori:2002uv}, which can be used to calculate self-force on Kerr spacetime.
\end{appendices}
\end{document}